\documentclass[numberedappendix,appendixfloats]{emulateapj}

\usepackage{graphicx} 
\usepackage{hyperref}	
\usepackage{subfigure}

\newcommand{\NII}{\hbox{[{\rm N}{\sc \,ii}] }}

\newcommand{\Ha}{\hbox{{\rm H}$\alpha$}}

\newcommand{\flux}{erg\,s$^{-1}$\,cm$^{-2}$}

\newcommand{\mpy}{\hbox{M$_{\odot}$\,yr$^{-1}$}}

\newcommand{\msun}{\hbox{M$_{\odot}$}}

\newcommand{\kms}{\hbox{${\rm km\,s}^{-1}$}}
\newcommand{\nn}{\nonumber}

\newcommand{\rmin}{1.5}

\newcommand{\be}{\begin{eqnarray}}
\newcommand{\ee}{\end{eqnarray}}

\newcommand{\galpak}{{GalPaK$^{\rm 3D}$}}
\newcommand{\Nideal}{$1728$} 
\newcommand{\sizes}{ 2.5,  5, and  7.5} 
\newcommand{\sizesArc}{0\farcs3, 0\farcs6, and 1\farcs0} 
\slugcomment{Received 2014 June 24; accepted 2015 June 16}

\shorttitle{\galpak: a 3D kinematic tool}
\shortauthors{Bouch\'e et al.}

\begin{document}
\title{\galpak: A Bayesian parametric tool for extracting morphokinematics of galaxies from 3D data}

\author{N.  Bouch\'e\altaffilmark{1,2},  
H. Carfantan\altaffilmark{1,2},  I. Schroetter\altaffilmark{1,2},  L. Michel-Dansac\altaffilmark{3},  T. Contini\altaffilmark{1,2}
}

\altaffiltext{1}{CNRS/IRAP, 14 Avenue E. Belin, F-31400 Toulouse, France}
\altaffiltext{2}{University Paul Sabatier of Toulouse/ UPS-OMP/ IRAP, F-31400 Toulouse, France }
\altaffiltext{3}{CRAL, Observatoire de Lyon, Universit\'e Lyon 1, 9 Avenue Ch. Andr\'e, 69561 Saint Genis Laval Cedex, France}
 

\begin{abstract}
We present a method to constrain galaxy parameters directly  from three-dimensional data cubes.
The algorithm   compares directly the data   with a parametric model mapped in  $x,y,\lambda$ coordinates.
It uses the spectral line-spread function (LSF) and the spatial point-spread function (PSF) to generate a 3-dimensional kernel 
whose characteristics  are instrument specific or  user generated.
The algorithm returns the intrinsic modeled properties along with both
an `intrinsic' model data cube and the modeled galaxy convolved with the 3D-kernel. 
The algorithm uses a Markov Chain Monte Carlo (MCMC) approach  with a  nontraditional
 proposal distribution in order to efficiently probe the parameter space.
We demonstrate the robustness of the algorithm using \Nideal\ mock galaxies and
 galaxies generated from hydrodynamical simulations   in various seeing conditions from 0\farcs6 to 1\farcs2.  
We find that the algorithm can recover  the morphological parameters (inclination, position angle) to within 10\%\ and the kinematic parameters (maximum rotation velocity) to within 20\%, irrespectively of the PSF in seeing (up to 1\farcs2) provided that the maximum signal-to-noise ratio (SNR)  is greater than $\sim3$ pixel$^{-1}$ and that ratio of galaxy half-light radius to seeing radius   is greater than about \rmin.
One can use such an algorithm  to constrain simultaneously the kinematics
and morphological parameters of (nonmerging) galaxies observed in nonoptimal seeing conditions.
The algorithm can also be used on  adaptive-optics (AO) data
or on high-quality, high-S/N data   to look for nonaxisymmetric structures in the residuals.
\end{abstract}

\section{Introduction}

Thanks to several studies using optical or near-infrared (NIR) integral field unit (IFU) spectroscopy of \Ha\ emission from local and high-redshift  ($z>1$) galaxies
 \citep{ForsterSchreiberN_06a,LawD_07a,VanstarkenburgL_08a,CresciG_09a,ForsterSchreiberN_09a,Lemoine-BusserolleM_10a,LawD_12a,ContiniT_12a,EpinatB_12a,BuitragoF_14a}, our understanding of galaxy formation has changed significantly in the past decade.
For instance, these surveys have shown that a significant subset of
high-redshift galaxies have a disklike morphology and show   organized rotation, with regular velocity fields.

In contrast to low-redshift studies \citep[e.g.][]{BaconR_01a,CappellariM_11a},   high-redshift ($1\lesssim z\lesssim2$) galaxies are observed at a   spatial resolution that is severaly  limited by the
seeing conditions  owing to their small apparent angular sizes.  
In order to overcome the low spatial resolution, observations with adaptive-optics (AO) are often required \citep{LawD_07a,LawD_09a,GenzelR_08a,GenzelR_11a,WrightS_07a,WrightS_09a}.
However, observations with AO are expensive, with the additional instrumental costs, and add strong observational constraints such as the additional exposure times required to compensate for the loss
in  surface brightness (SB) sensitivity \citep{LawD_06a}. Indeed, the SB limit for AO observations taken on smaller pixels is higher, leaving
the current state-of-the art observations to the objects with the highest 
SBs.

Given these challenges and the advancements in multiplexing IFU observations with 
the Very Large Telescope (VLT) second-generation instruments like KMOS \citep{SharplesR_06a} and the Multi-unit Spectrograph Explorer \citep[MUSE;][]{BaconR_06a,BaconR_15a},
it is important to have tools that can give robust estimates on the galaxy physical properties.
 In particular, KMOS will bring large statistically significant samples of high-redshift galaxies as it can observe 24 galaxies at a time, but this facility will always lack an AO unit. 
 This could potentially be a serious limitation
since the robustness of the derived kinematic parameters may depend on the quality of the atmospheric conditions (seeing can range from 0\farcs4 to $>$1\farcs0 in the NIR).

In order to overcome these limitations,
we present  a new tool named \galpak\ (Galaxy Parameters and Kinematics)~\footnote{Available at \url{http://galpak.irap.omp.eu/}.} designed
to be able to disentangle the galaxy kinematics from resolution effects over a wide range of conditions.  
This is not the first code to model galaxy kinematics  from 3D data \citep[e.g. the TiRiFiC package,
which performs tilted ring model fits to three-dimensional radio data;][]{JozsaG_07a},
 but this code performs disk model fits to three-dimensional
IFU data cubes, for the first time,~\footnote{\citet{LawD_12a} made an attempt at 3D fitting, albeit not self-consistently.},
whereas all other modeling of IFU-data so far have worked from the two-dimensional velocity field
 \citep[e.g.][]{CresciG_09a,EpinatB_09a,DaviesR_11a,AndersenD_13a,DavisT_13a}.

This paper is organized as follows:   we describe  the GalPaK$^{3D}$ algorithm  in Section~\ref{section:galpak}.
We present some test case examples in Section~\ref{section:highlights}.
We present results from an extensive analysis of \Nideal\  synthetic galaxies in Section~\ref{section:simu},  
where we discuss the impact of the accuracy in the point-spread-function (PSF) characterization.
 In Section~\ref{section:Leo}, we present an analysis of data cubes generated from hydrodynamical simulations of isolated
disks from Michel-Dansac et al. (in prep.).  We summarize this paper in Section~\ref{section:conclusions}. 
Throughout, we use the following cosmological parameters:  $H_0=70$~km~s$^{-1}$, $\Omega_{\Lambda}=0.7$ and $\Omega_{M}=0.3$.


\section{The GalPak$^{3D}$ algorithm}
\label{section:galpak}

In this section we outline the algorithm principles, which are designed to be able to determine galaxy morphokinematic parameters from the three-dimensional data cube directly.
We discuss the merits of using the parametric forward fit and its limitations.

\subsection{A Parametric Galaxy Model in Three Dimensions}

Traditionally, kinematic analyses use two-dimensional
 maps generated by applying line-fitting codes to determine the line
 wavelength centroids and widths, which are only considered to be
 reliable for spaxels with sufficiently high signal-to-noise (S/N) ratios.
This S/N condition is easily met at low redshifts, but is harder to meet for small, high-redshift galaxies.
In principle, the choice to work in 2D or 3D space is equivalent,
but we will show that our method can work in the regime (on the spaxels) where the signal-to-noise ratio (SNR) per pixel (SNR pixel$^{-1}$) is not sufficient for line-fitting codes, which require a minimum SNR on all spaxels.

When the PSF FWHM can be characterized to sufficient accuracy~\footnote{The PSF shape matters more than the level of accuracy on the FWHM, as discussed in Section~\ref{section:PSF}.}
 (within 10\%\ or 20\%; see Section~\ref{section:simu}),  
one can take its characteristics, together with the instrumental line-spread function (LSF), into consideration and recover the intrinsic modeled galaxy parameters.
The algorithm uses the spectral LSF and the spatial PSF  to generate a three-dimensional kernel 
whose characteristics are set for the given instrument (or a user-generated  instrument module).

While a full deconvolution of hyperspectral cubes would be preferred, it is usually a challenge mathematically 
\citep[a new method has been proposed recently by][]{VilleneuveE_14a}, and a forward convolution of a parametric model offers
a very useful alternative.  This forward convolution  gives us the opportunity to estimate intrinsic modeled kinematic parameters in a wider range of seeing conditions,
as illustrated in  recent papers \citep[see][for first applications]{BoucheN_13a,PerouxC_14a,SchroetterI_15a,SotoK_15a}.

For the forward convolution, we need a parametric model, and we focus here on a galaxy disk model for emission-line surveys, but the algorithm is adaptable to other situations.
In order to construct a modeled galaxy in the observational coordinate
systems ($x$, $y$, $\lambda$), we start by generating a three-dimensional galaxy model in a Euclidian coordinate system ($x$, $y$, $z$), where the $z$-axis is normal to the galaxy plane ($x$, $y$). 
We apply a  radial flux profile $I(r)$, from one of the traditional
Gaussian, exponential, and de Vaucouleur choices as parameterized by the \citet{SersicJ_63a} profile:
\begin{eqnarray}
I(r)&=&I_{\rm e} \,\exp\left(-b_n\,\left[(r/R_{\rm e})^{1/n}-1\right]\right) \label{eq:Ir} 
\end{eqnarray}
with $n=0.5$, 1.0, and 4.0, respectively,
where $R_{\rm e}$ is the effictive radius, $f_{\rm tot}$ the total flux, and $b_n$ such that $R_{\rm e}$   is equivalent to the half-light radius $R_{1/2}$,
and $I_{\rm e}$ the SB at $R_{\rm e}$.
For $n=0.5$, 1.0, and 4.0, the constant $b_n$ is 0.69, 1.68, and 7.67, respectively, from $b_n\simeq1.9992\,n-0.3271$.
The Sersic index $n$ is kept fixed given the large degeneracies it creates with other parameters, such as the galaxy half-light radius.
This degenaracy is 
due to the fact that the SB profiles around $R_{\rm e}$ are close to one another fo $n=0.5$, 1.0, or 4.0 as noted in \citet{GrahamA_05a}.

To this two-dimensional disk model, we add a disk thickness $h_z$. 
We adopt a Gaussian luminosity distribution
 perpendicular to the plane,   $I(z) \propto \exp(-z^2/2\,h_z^2)$, defining $h_z$
 as the characteristic thickness of the disk. 
\galpak\  also allows the user to choose an exponential $I(z) \propto\exp(-|z|/h_z)$ or a sech$^2$ distribution $I(z) \propto\;$sech$^2(z/h_z)$.
We set the disk thickness to $h_z=0.15\,R_{1/2}$ where $R_{1/2}$ is the disk half-light radius.
This choice corresponds to $h_z\sim1$~kpc, typical of high-redshift edge-on/chain galaxies \citep{ElmegreenB_06a}.
At this stage, we have a disk model  in Euclidean coordinates that accounts for the flux distribution only.

For the gas kinematics, we create three kinematic cubes in the same spatial coordinate reference frame for the velocities
$\mathbf {v}=(V_x,V_y,V_z)$ assuming  circular orbits. The rotational velocity $v(r)$ with 
 a maximum rotation velocity $V_{\rm max}$ can have several functional forms:
it can be an $\arctan$ velocity profile \citep[e.g.][]{PuechM_08a},
an inverted exponential \citep{FengJ_11a}, or 
a hyperbolic $\tanh$ profile \citep[e.g.][]{AndersenD_13a} :
\begin{eqnarray}
v(r)&=&V_{\rm max}\frac{2}{\pi}\arctan\left(r/r_{\rm t} \right) \qquad\hbox{`arctan'}\\
v(r)&=&V_{\rm max}\,\left[1-\exp(r/r_{\rm t})\right]\qquad\hbox{`exp'}\\
v(r)&=&V_{\rm max}\,\tanh(r/r_{\rm t})\qquad\hbox{`tanh'}
\end{eqnarray}
where $r$ is the radius in the galaxy $x,y$ plane, $r_{\rm t}$ is the turnover radius, and $V_{\rm max}$ is the maximum circular velocity. 
 These choices are more extensively discussed in \citet{EpinatB_10a},
but it is worth noting that the `exp' and hyperbolic rotation curves have a sharper transition around the turnover radius.
We stress that our parameter $V_{\rm max}$ is {\it not} the projected asymptotic velocity, but is the true asymptotic velocity irrespective of the inclination.

 Another option, called ``mass,'' assumes a constant light-to-mass ratio and sets $v(r)$ from the
enclosed light/mass $I(<r)$ profile
\begin{eqnarray}
v(r)&\propto& \sqrt{\frac{I(<r)}{r}}\qquad\hbox{`mass'}
\end{eqnarray}
where $r$ is the radius in the galaxy $x,y$ plane and $V_{\rm max}$ normalizes the profile.
This option has a  rotation curve that peaks at some radius (set by the half-light radius),
 decreases at larger radii, and is to be preferred for nuclear disks or when there is no significant dark matter component.

We then rotate the disk model around two axes according to an inclination ($i$) and position angle (PA, anticlockwise from $y$) and create a cube in $x$, $y$, and $\lambda$
using the three intermediate 2D maps: the flux map, the velocity field, and the dispersion map ($\sigma_{\rm tot}$). 
The flux map is obtained from the rotated flux cube summed along the wavelength axis.
The velocity field is obtained from the flux-weighted mean $V_z$ velocity cube.
The total (line-of-sight) velocity dispersion $\sigma_{\rm tot}$ is obtained from the sum of three  terms (added in quadrature).
It includes (i) the local isotropic velocity dispersion $\sigma_{\rm d}$ driven by the disk self-gravity, which is $\sigma_{\rm d}(r)/h_z=V(r)/r$ 
for a compact ``thick''  or large ``thin''  disk \citep{GenzelR_08a, BinneyJ_08a,DaviesR_11a};
(ii) a mixing term,  $\sigma_{\rm m}$, arising from mixing the velocities along the line of sight for a geometrically thick  disk,
which is obtained from the flux-weighted variance of the cube $V_z$,
and (iii)  an intrinsic  dispersion ($\sigma_o$)  ---which is assumed to be isotropic and constant spatially---  
to account for the fact that high-redshift disks are dynamically hotter than the self-gravity expectation.
Indeed, this turbulence term $\sigma_o$ is  often observed to be $\simeq50$--80\kms\  in  $z>1$ disks \citep{LawD_07a,LawD_12a,GenzelR_08a,CresciG_09a,ForsterSchreiberN_09a,WrightS_09a,EpinatB_10a,EpinatB_12a,WisnioskiE_11a}
 and thus dominates the other two terms since the mixing term $\sigma_{\rm m}$  is typically $\sim$15 \kms\
 and the self-gravity term $\sigma_{\rm d}$ is typically 10--30\kms.

To summarize, the flux profile can be chosen to be `exponential' ($n=1.0$),`gaussian' ($n=0.5$), and `de Vaucouleur' ($n=4.0$); the velocity profile $v(r)$ can be arctan (``arctan''), inverted exponential (``exponential''), hyperbolic (``tanh'') or that of mass profile (``mass'');
and the local dispersion can be that of the thin or thick disk.  
There are in total 10 free parameters~\footnote{There are only nine free parameters when the ``mass'' profile is used for $v(r)$ since the turnover radius $r_{\rm t}$ is irrelevant.}
 to be determined from the data.
The 10 parameters are the $x_{\rm c}$, $y_{\rm c}$, $z_{\rm c}$ positions, the disk half-light radius $R_{1/2}$, the total flux $f_{\rm tot}$, the inclination $i$, position angle PA, the turnover radius $r_{\rm t}$, the maximum circular velocity $V_{\rm max}$,
and the one-dimensional intrinsic dispersion $\sigma_o$ .
We will refer to the last two ($V_{\rm max}$, $\sigma_o$) as kinematic parameters.
Finally, the simulated galaxy is convolved (in 3D) with the PSF and the instrumental LSF specific for each instrument.\footnote{The user can choose a Gaussian PSF, a Moffat PSF. The PSF can be circular or elliptical with a user-defined axis ratio $b/a$.}
The 3D convolution is performed using fast Fourier transform (FFT) libraries.


\subsection{The Markov Chain Monte Carlo (MCMC) Algorithm}

\begin{table}
\caption{Default Range on Each Parameter \label{table:boundaries}}
\centering
\begin{tabular}{lrr}	
Parameter & Min & Max \\
\hline
$x_{\rm c}$ & 1/3 $N$pix$_x$ & 2/3 $N$pix$_x$ \\
$y_{\rm c}$ & 1/3 $N$pix$_y$ & 2/3 $N$pix$_y$ \\
$z_{\rm c}$ & 1/3 $N$pix$_z$ & 2/3 $N$pix$_z$ \\
Flux   & 0 &  $3\times \Sigma_{i,j,k}(v_{i,j,k})$  \\
$R_{1/2}$ & 0.2 spaxel & 4 \arcsec\\
Incl. ($\deg$) & $0$ & $90 $ \\
PA ($\deg$) & $-180$ & $180 $\\
$r_{\rm t}$ & 0.01 spaxel & 1 \arcsec\\
$V_{\rm max} ( \kms) $ & -350 & 350   \\
$\sigma_o$   ( \kms) & 0  & 180 
\end{tabular}
\end{table}

In order to determine the 10 free parameters on hyperspectral cubes, one needs an algorithm
that is independent of initial guesses on the parameters, that can converge even in the presence of local minima, and that can handle low S/N data.
This is particularly difficult  for traditional minimization methods 
 because the $\chi^2$ hypersurface is very flat (outside the shallow well near the optimum parameters), and as a result 
the minimization algorithm tends to not converge and be very susceptible to local minima.  

Here  we use an algorithm  to optimize the parameters using Bayesian statistics with flat priors on bound intervals for each of the parameters. 
The  algorithm constructs MCMCs  with a Metropolis-Hasting (MH) sampler \citep{HastingsW_70a}. 
At each iteration we compute the new set of parameters $\hat x_{i+1}$ from the last $\hat x_{i}$ set with a proposal distribution $P$ from which to draw:
\begin{equation}
\hat x_{i+1}=\hat x_{i}+\hat h\, P(\hat x_{i+1}|\hat x_i),\label{eq:P}
\end{equation}
where the new set of parameters is accepted or rejected as in any MH algorithm.  
The new proposal set of parameters $x_{i+1}$ is then accepted or rejected 
according to the posterior distribution, which amounts to the likelihood ${\cal L}\propto\exp -\chi^2$ in the considered case of flat priors on the parameters.
In other words, we assume that the pixels are independent and that noise properties are Gaussian, which is appropriate for optical/NIR data taken in the background-limited regime,
and the user can provide the full variance cube.
More appropriate likelihood functions for low counts with  Poisson noise can be found in \citet{MighellK_99a}.

The scaling vector $\hat h$ in Equation~\ref{eq:P} is derived from the variance on the flat (uniform) prior distributions, whose boundaries are adjustable
(the default values are   listed in Table~\ref{table:boundaries}).
The user  may need to  rescale the vector $\hat h$ in order to have acceptance rates between 20\%\ and 50\%.
Convergence is usually achieved in a few hundred to a few thousand iterations, even
though we typically let the algorithm run for 15,000 iterations. 

In principle, one has the freedom to use {\it any} proposal distribution $P$  \citep[e.g.][]{MacKayD_03a}.
A Markov chain is said to converge to a single invariant distribution (the posterior probability) when the state of the chain persits once it is reached
and is said to be ergodic when the probabilities $x_n$ converge to that invariant distribution as $n\rightarrow\infty$, irrespectively of the initial parameters
\citep{NealR_93a}.\footnote{Available at \url{http://www.cs.toronto.edu/~radford/res-mcmc.html}.}
In addition, if the sampler satisfies the following   conditions  $P(x|x')=P(x'|x)$,  as we have used, the algorithm reverts to the Metropolis method,
which satisfies the two conditions.
In practice, however, one also needs a distribution that probes the parameter space efficiently in order to
 converge in a reasonable number of cpu hours, regardless of the initial parameters.

A common proposal distribution is the uniform distribution that gives equal probabilities to all possible values.
The Gaussian proposal distribution $P(x'|x)= {\cal N}(x,1)$ is probably the most commonly used and is popular but has one major drawback: the Gaussian distribution is rather narrow such that the algorithm becomes sensitive to the initial conditions, making the time to convergence to the optimum values very sensitive to the initial guess. 
If the width of the proposal distribution is small, the convergence is too slow/large, and when it is large (for convergence purposes), it will lead to low acceptance rate and poor efficiencies for convergence. 
To remedy this problem, one could use a mixed distribution with a Gaussian draw, say, 90\%\ of the time and a uniform draw 10\%\ of the time, allowing
the chain to escape from a local minimum.  Compared to the Gaussian proposal, the mixed
distribution has one additional parameter that needs to be fine-tuned to the problem, such as the  
mixing ratio.

A third option, as advocated by \citet{Szu_87a}, is to use a draw from a Cauchy distribution
that has by definition longer wings (i.e. $P$ is a Lorentzian profile where $P(x'|x)\propto\gamma^2/[\gamma^2+(x'-x)^2]$).
The Lorentzian wings are important, allowing
 the chain to make large jumps during the initial ``burn-in'' phase and
 ensuring rapid convergence of the chain with no sensitivity to the
 initial parameters.
Another advantage of a Cauchy proposal distribution is that it has only one parameter, $\gamma$, compared to the mixed one.

 We tested these various choices on simulated cubes and  found that the Cauchy proposal distribution converged faster than the other methods and was least sensitive to the initial parameters.   
In other words, with the Cauchy   nontraditional proposal distribution,  a few hundred to a few thousand steps of the MCMC are required to pass the  burn-in phase depending on the S/N of the data,
 and it is the user's responsibility to confirm that the MCMC chain has converged. 
Thus, we typically run the chain  through 10,000 or 15,000 steps to robustly sample the posterior probability distribution.

The ``best-fit'' values of the parameters are determined from the posterior distributions. 
We use the median and the standard deviation of the last fraction (default 60\%) of the MCMC chain
to determine the `best-fit' parameters and their errors, respectively.   
One can also use a fraction (default 60\%) of the MCMC chain around the minimum $\chi^2$.
The full MCMC chain is saved such that the user can use his/her preferred technique.

The algorithm is implemented in Python and uses the standard numpy and scipy libraires.
In addition, it uses the bottleneck~\footnote{Available at \url{https://pypi.python.org/pypi/Bottleneck}.} \citep{FrigoM_05a} and  FFTw~\footnote{Available at \url{https://pypi.python.org/pypi/pyFFTW}.} libraries \citep{FrigoM_12a} in order to speed up certain matrix 
operations and the PSF$+$LSF convolution, respectively. It requires FITS files as inputs.  
The algorithm is modular so that the user can add specifications for other instruments. 
The online documentation describes the syntax, and it takes about 2, 5, and 10 minutes on a laptop  (at 2.1 GHz) 
to run 10,000 iterations on a data cube with 30$^3$ pixels, 40$^3$ pixels, and 60$^3$ pixels, respectively.
In other words, the computation time scales  as $t\propto N_{\rm pix}\log(N_{\rm pix})$ where  $N_{\rm pix}$  is the number of pixels, showing that the FFT calculation dominates.


\section{Highlight applications}
\label{section:highlights}

\begin{figure*}
\centering
\includegraphics[width=16cm]{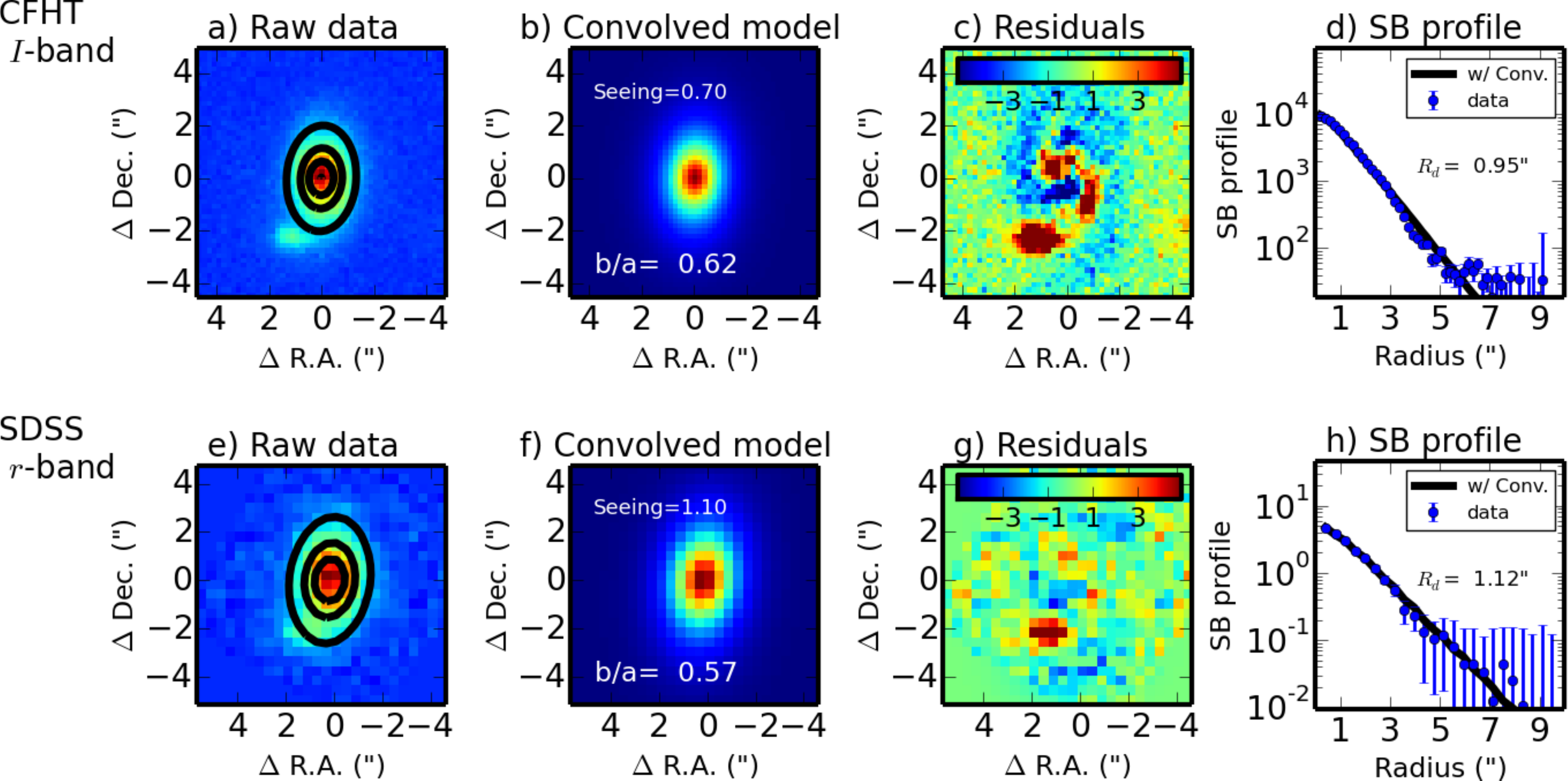}
\caption{Application of the two-dimensional version of the MCMC algorithm (`GalFit{2D}') on the  $z\sim0.2$  SDSSJ165931.92$+$023021.9
with $m_r=$18.40~mag from  \citet{KacprzakG_14a}.
Similarly to \galpak, Galfit2D performs a parametric fit with an MCMC algorithm using set surface brightness profiles convolved with the seeing.
The top row shows the result from archival CFHT $I$-band taken at a resolution of 0\farcs7.
The bottom row shows the result from the SDSS $r$-band image that has a resolution of 1\farcs1.
Panels (a) and (e) show the data.
Using an exponential profile, panels (b) and (f) show the seeing-convolved model;  (c) and (g) the residuals, i.e. data-model normalized to the pixel noise $\sigma$, and
(d) \&\ (h) the one-dimensional SB profile. The recovered intrinsic disk scale length $R_{\rm d}$ is about 1" in both cases, in spite of the different spatial resolution.
\label{fig:J1659}}
\end{figure*}

\begin{figure*}
  \centering
 \includegraphics[width=18cm]{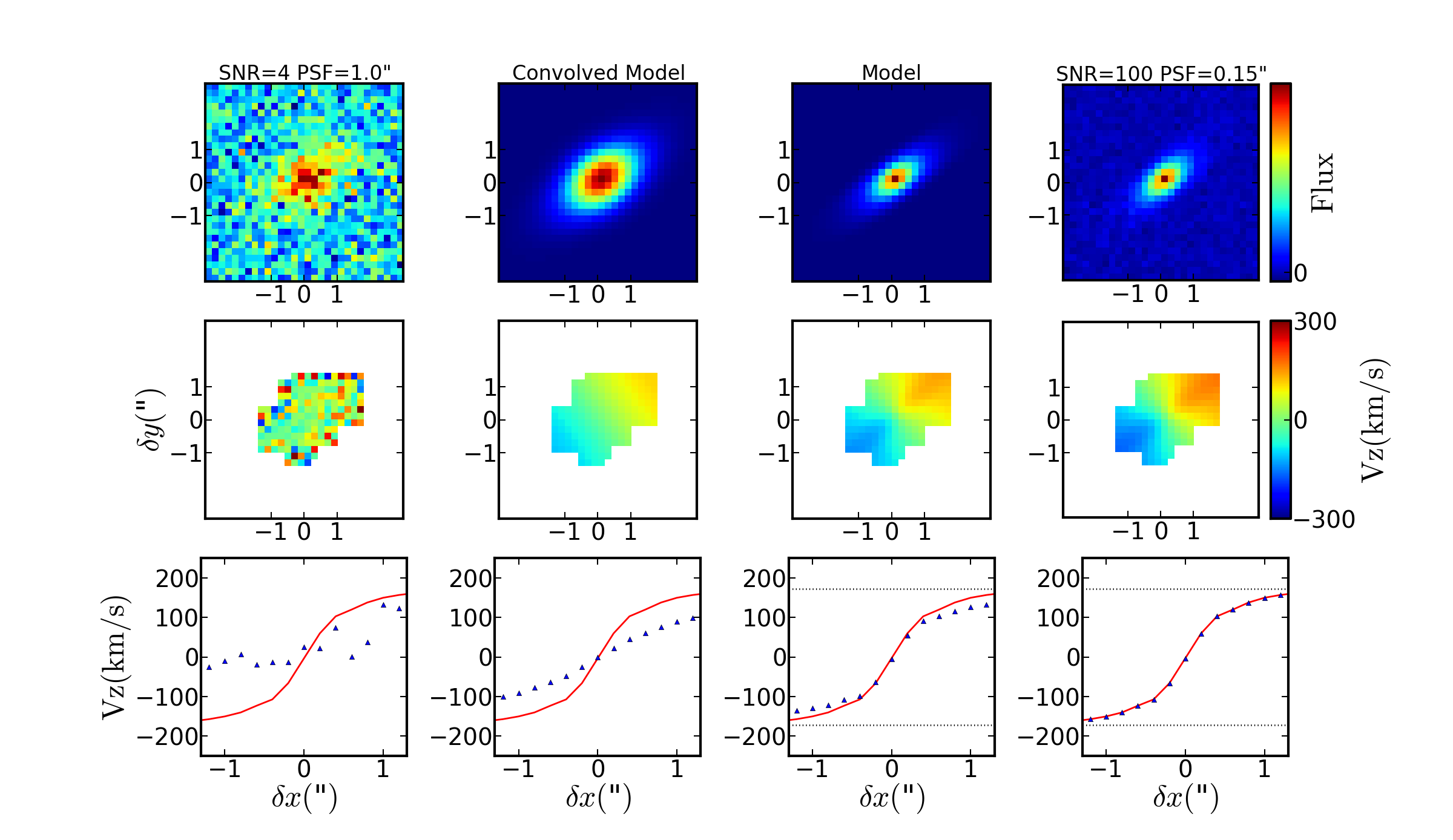}  
  \caption{Example of the algorithm application on a disk model simulated with  a seeing of 1\farcs0 (FWHM) and a flux of
$10^{-16}$~\flux, and  an  S/N pixel$^{-1}$ of $\sim$4 {at the brightest pixel}.  
The top, middle, and bottom rows show the  flux map, the velocity map,  and the apparent velocity profile $V_z(r)$ across the major axis, respectively.
From left to right, the panel columns show the data, the convolved model, the modeled disk (free from the PSF),
 and the high-S/N high-resolution reference data (PSF=0\farcs15 and SNR=100). 
 In the bottom panels, the solid red curves  correspond to the reference case, the triangles represent the apparent rotation curve, and  the dotted lines show the apparent  $V_{\rm max}\,\sin i$.
One sees that the velocity profile from the modeled disk (third column) is in good agreement with the reference data.}
\label{fig:SNR3} 
\end{figure*}

\subsection{Example on 2D data}

Before applying the tool on 3D data, it is important to validate the method on simpler data sets, such as two-dimensional imaging data.
We thus wrote a two-dimensional version of the algorithm, GalFit$^{2D}$, one that does not include the kinematic, which is in essence similar to other parametric algorithms \citep[e.g.][]{SimardL_98a,PengC_02a}, apart from the Bayesian approach.

Figure~\ref{fig:J1659} shows a comparison between the derived morphological parameters from two data sets of very different resolution.
Panel (a) shows a  Canada France Hawaii Telescope (CFHT) $I$-band image of the $z\sim0.2$ galaxy SDSSJ165931.92$+$023021.92 \citep{KacprzakG_14a}.
Panel (e) shows an $r$-band image of the galaxy from the Sloan Digital Sky Survey (SDSS) at a spatial resolution of 1\farcs1.
For each data set, we show the fitted (convolved) model, the residual map, and the one-dimentional SB profile.
One sees that the intrinsic modeled morphological parameters found from the SDSS data (PSF FWHM=1\farcs1) are in good agreement with the higher-resolution data (PSF FWHM=0\farcs7). 
Moreover, the residuals in both data sets show the spiral arms and a minor merger (or a large clump) in the southern part of the galaxy,
showing that a smooth axis-symmetric model can be used to unveil asymmetric features.

\subsection{Example  on a mock cube }

Figure~\ref{fig:SNR3} shows an example of a mock disk model with a low SNR (SNR pixel$^{-1}$ of 4 in the central pixel)
drawn from the set presented in \S~\ref{section:simu} and generated at 1\farcs0 resolution.
The top, middle, and bottom rows show the  flux map, the velocity map, and the apparent velocity profile $V_z(r)$ across the major axis, respectively.
From left to right, the panel columns show the data, the convolved model, the modeled disk (free from the PSF),
 and the high-S/N high-resolution reference data (PSF=0\farcs15 and S/N=100). 
 In the bottom panels, the solid red curves  correspond to the reference rotation curve (obtained from the reference data set),
and 
the triangles represent the apparent rotation curve.
These rotation curves show that the recovered kinematics from the modeled disk (intrinsic or unconvolved model) shown in the third column
 is in good agreement with the reference data (last column) in spite of the low spatial resolution (1\farcs0) and the low SNR in the mock data set.

\begin{figure}
\centering
\subfigure[]{
\includegraphics[width=5cm]{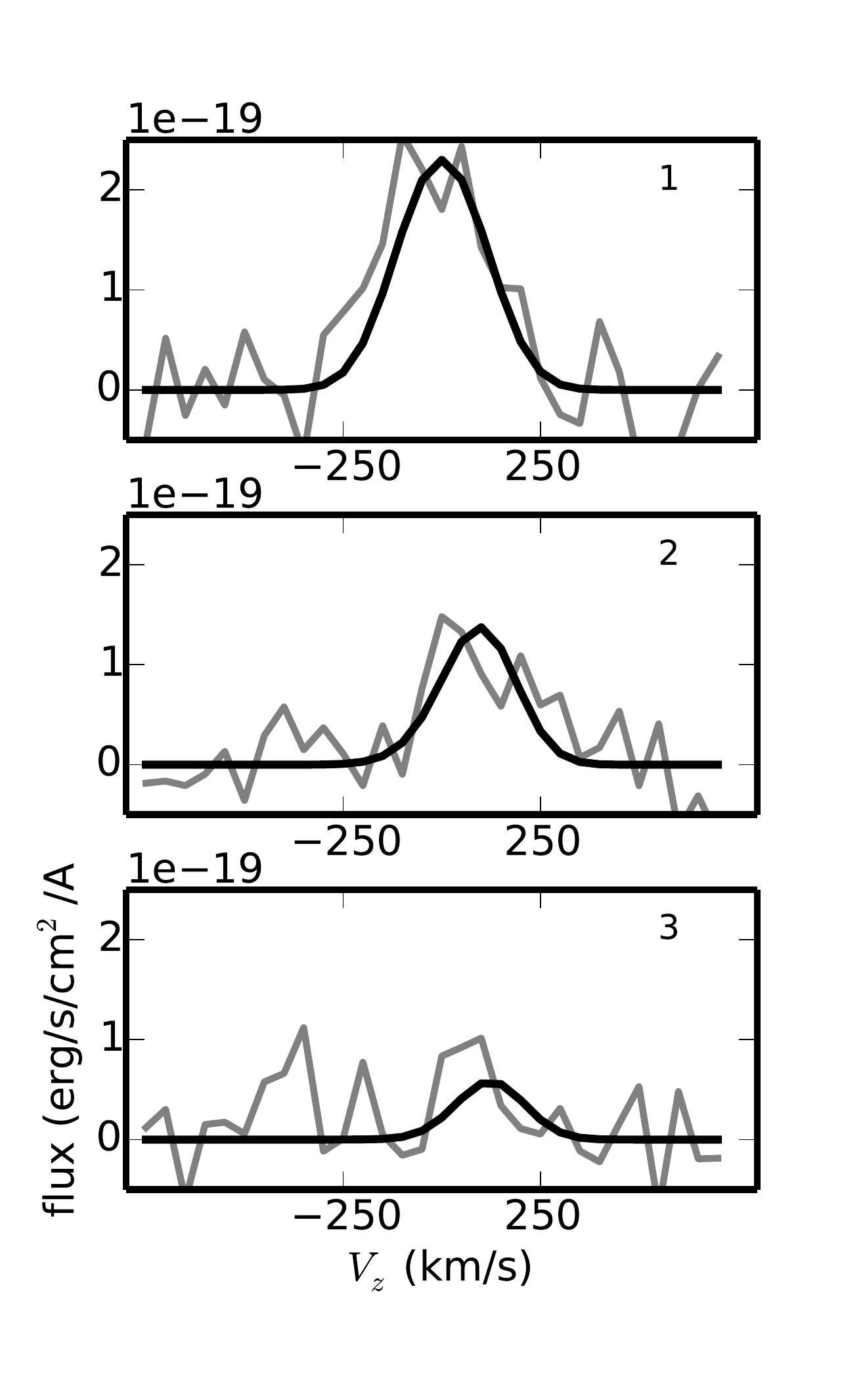}
}
\subfigure[]{
\includegraphics[width=6cm]{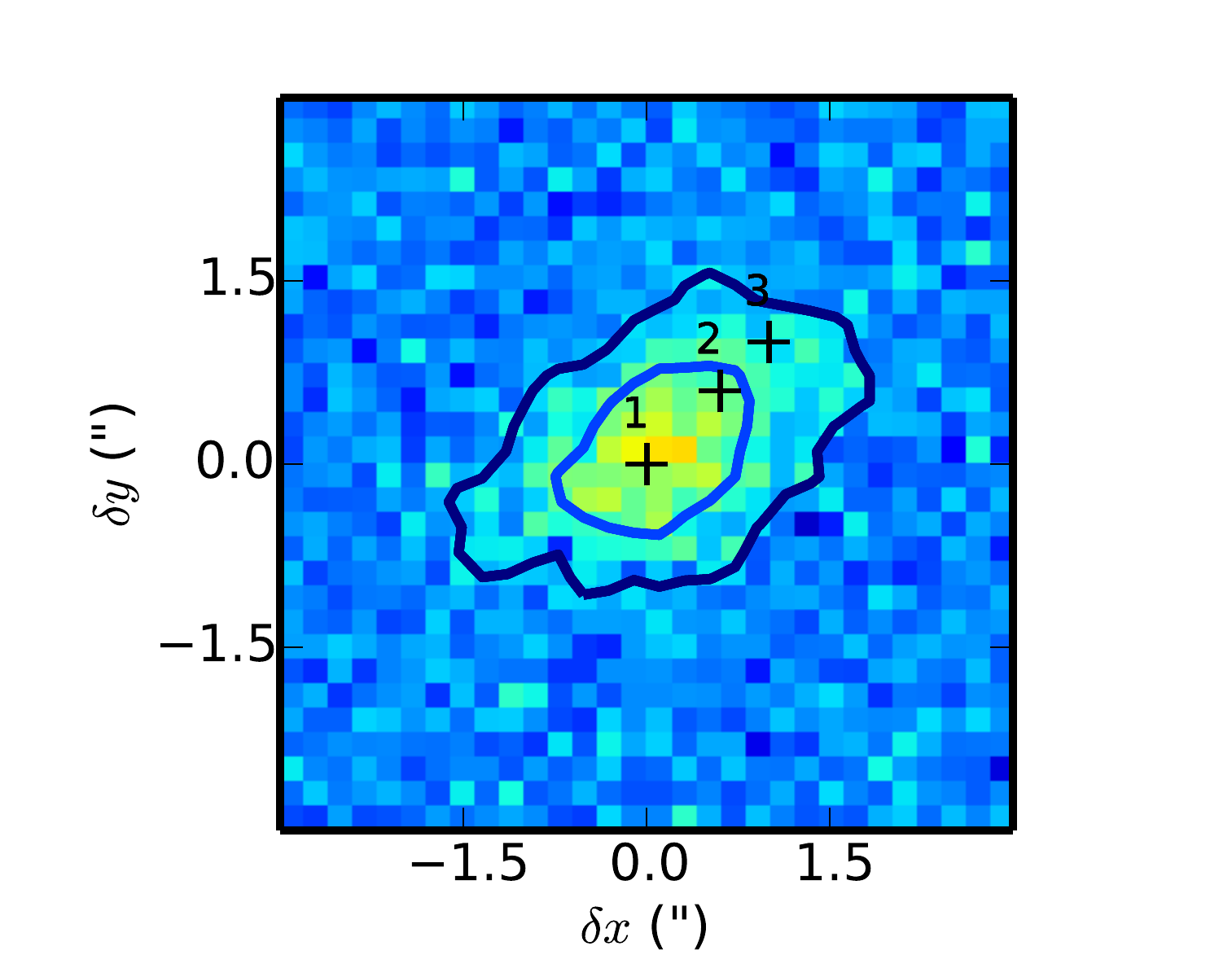}
}
\subfigure[]{
\includegraphics[width=6cm]{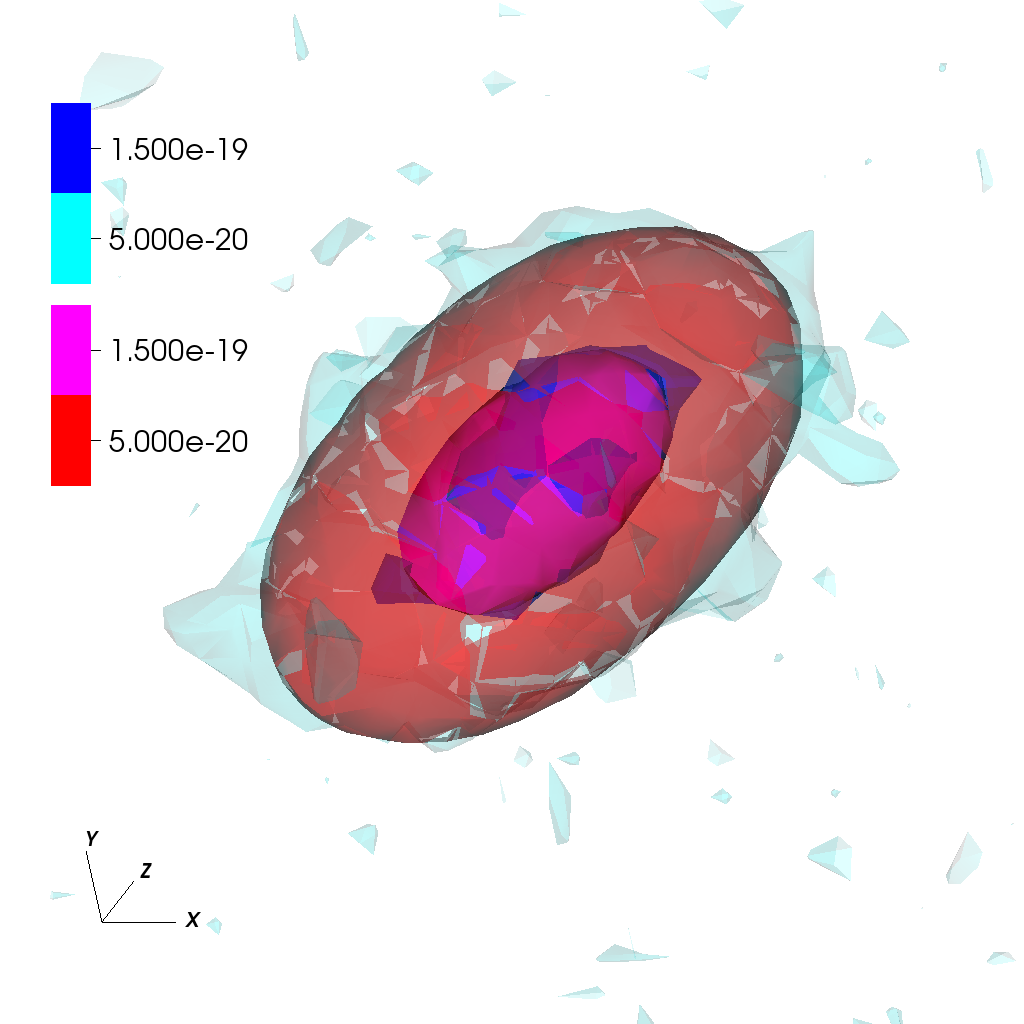}
}
\caption{For the synthetic data  in Figure~\ref{fig:SNR3}, we show three 1-dimensional profiles (panels (a)) comparing the data (thin line) and the model (thick line) taken at the location
labeled ``1,'', ``2,'' and ``3'' in panel  (b).  Panel (c) shows a 3D representation of the data (blue) with the model overlaid (red) where we
used two flux levels at  5$\times10^{-20}$ and 1.5$\times10^{-19}$~\flux~\AA$^{-1}$. The cube orientation is shown, where the wavelength axis is the $z$-direction.
(An interactive version of this figure is available in the published online version) \label{fig:137_3D}  }
\end{figure}

 This synthetic data cube was generated with a  flux profile with Sersic index $n=1$ and half-light radius $R_{1/2}=0\farcs5$, corresponding to  2.5~MUSE/KMOS pixels), an ``arctan'' velocity profile with $V_{\rm max}=200$~\kms, a thick disk with a velocity dispersion $\sigma_o=80$~\kms,  an inclination  $i=60^\circ$, a PA$=130^\circ$, and with instrumental specifications for the new VLT MUSE instrument (0\farcs2 pixel$^{-1}$, 1.25\AA pixel$^{-1}$, LSF=2.14 pixels).
The integrated total flux is $10^{-16}$~\flux, and the synthetic noise per pixel is $\sigma = 5\times10^{-20}$~\flux~\AA$^{-1}$.

The synthetic data cube is also displayed in 
Figure~\ref{fig:137_3D}, which shows three one-dimensional  spectra (a)  taken at the
 three locations labeled in the image shown in panel (b). 
Panel (c) shows a 3D representation of the data (blue) with the model overlaid (red) made with the ``visit'' software;\footnote{Available at \url{http://visit.llnl.org/}.}
where
 the light/dark areas corresponds to two cuts at fluxes of 6 and 8$\times10^{-20}$~\flux~\AA$^{-1}$, i.e. 
an S/N pixel$^{-1}$ of 1.2 and 1.8, respectively.

\begin{figure*}
\centering
\includegraphics[width=16cm]{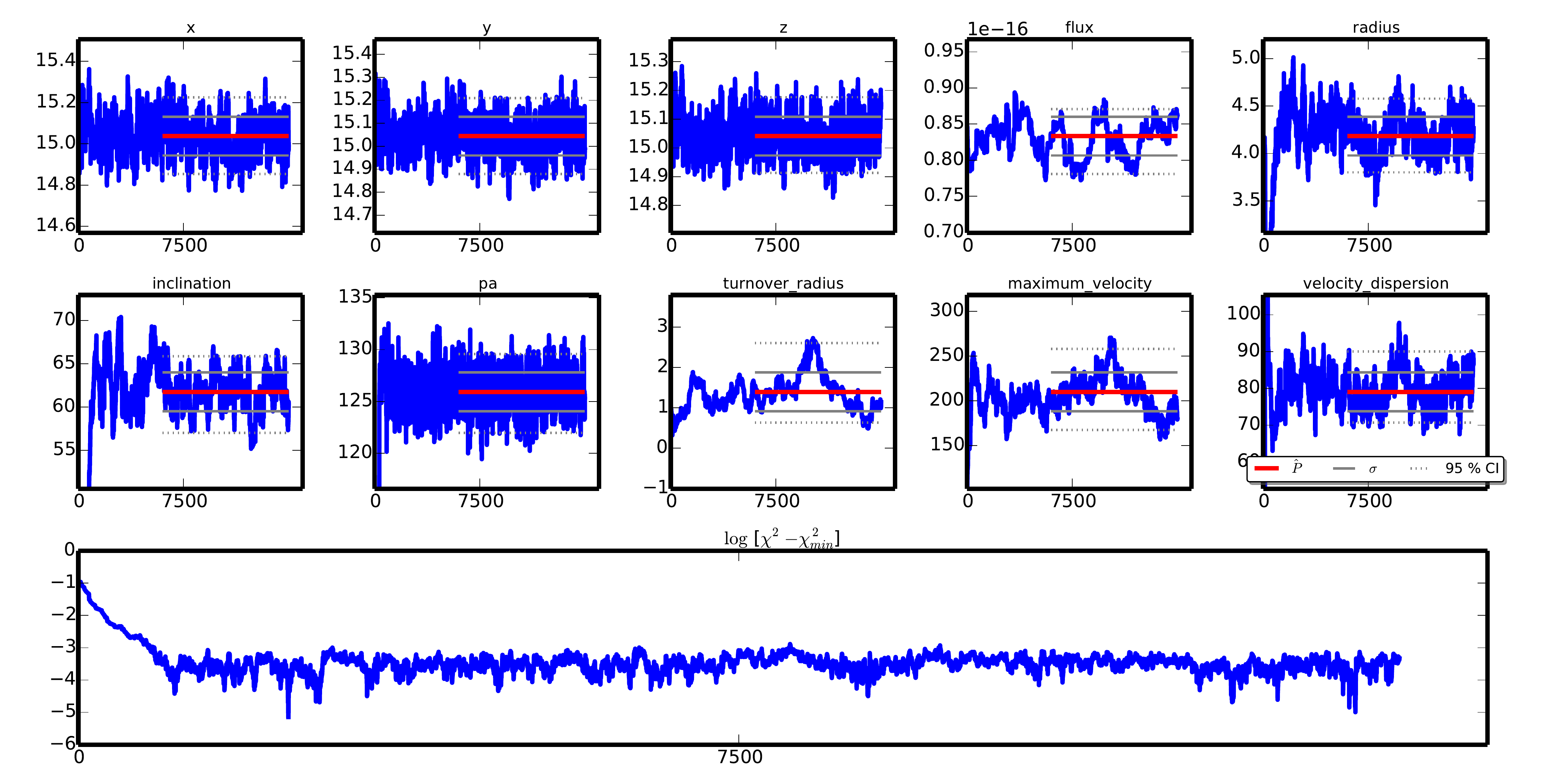} 
\caption{Full MCMC chain for 15,000 iterations for the example shown in Figure~\ref{fig:SNR3}. Each of the small panels corresponds to one parameter.
One sees that the ``burn-in'' region is confined to the first 1000 iterations. The estimated parameters are shown with the red line and are calculated from the last 60\%\ of the chain. The gray lines show the $1~\sigma$ standard deviations, and the dotted lines show the 95\%\ confidence interval.
 \label{fig:mcmc}
Note the fitted flux value is $(1.06\pm0.03)\times10^{-16}$~\flux, which is found from the sum of the pixel values (here $\sim 8.5\;10^{-17}$~\flux~\AA$^{-1}$) times the 1.25~\AA\ per spectral pixel.
 The bottom panel shows the $\chi^2$ evolution relative to the minimum, $\log [\chi^2-\chi^2_{\rm min}]$.
{We use this non-standard metric in order to show that the variations of the $\chi^2$ around the minimum which are 3 to 4
orders of magnitude smaller, reflecting a very flat hypersurface. Hence, a plot of $\chi^2$ or of the likelihood would show a straight line.}}
\end{figure*}

	We ran the algorithm with 15,000 iterations,  and Figure~\ref{fig:mcmc} shows the MCMC chains for the 10 free parameters along with the $\chi^2$ evolution  in the bottom panel. 
The values of the fitted parameters (and their errors) shown by the black lines (gray lines)
 are computed from the median (standard deviation) of the last 60\%\ iterations of  the posterior distributions. 
The recovered parameters are listed in Table~\ref{table:example}  and show  good agreement between the input   and   recovered values.

\begin{table}[ht]
\caption{Comparison between the model input values and the recovered values with 1~$\sigma$ errors and confidence intervals (CI) for the example shown in Figure~\ref{fig:SNR3}.
\label{table:example}} 
\centering
\begin{tabular}{lcc}
Parameter & Input & Output\hspace{0.5cm} [95\% CI]\\
\hline\\
$x_{\rm c}$ (pixel)& 15 & 15.05$\pm$0.09	\hfill [14.87;15.24]\\
$y_{\rm c}$ (pixel)& 15 & 15.06$\pm$0.09 	\hfill [14.89;15.23]\\
$z_{\rm c}$ (pixel)& 15 & 15.05$\pm$0.07 	\hfill [14.92;15.19]\\
Flux  ($10^{-16}$) & 1 & 1.06$\pm$0.03  \hfill [1.01;1.09]\\
$R_{1/2}$  (arcsec) & 0.82 & 0.85$\pm$0.04 \hfill  [0.78;0.95]\\ 
Incl. (deg) & 60 & 62$\pm$3  \hfill	 [58;68]\\
PA. (deg) & 130 & 126$\pm$2 \hfill	 [123;130] \\
$r_{\rm t}$ (pixel) & 1.35 & 1.32$\pm$0.42  \hfill  [0.8;2.47] \\
$V_{\rm max}$ (\kms) & 200 & 202$\pm$22  \hfill  [172;257] \\
 $\sigma_o$ (\kms)&  80 & 82$\pm$5  	\hfill  [73;90] \\
\hline
\end{tabular}
\end{table}

\begin{figure*}
\centering
\includegraphics[width=16cm]{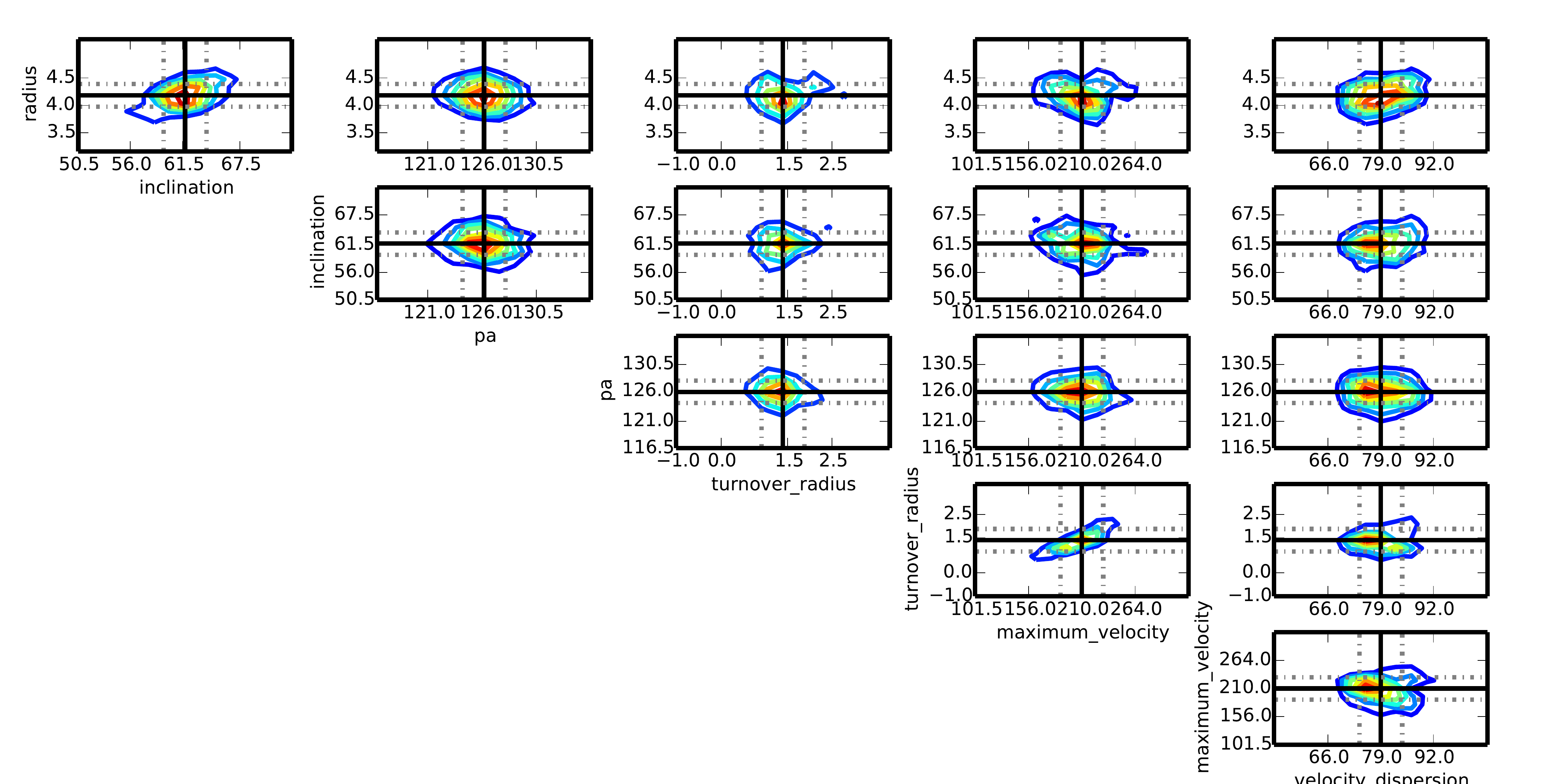} 
\caption{Joint distributions for the radius, PA, inclination, maximum velocity, and dispersion parameters for the example shown in Figure~\ref{fig:SNR3}--\ref{fig:mcmc}. The estimated parameters and their respective $1~\sigma$ error are shown as
a solid line and dashed line, respectively. One sees that the traditional degeneracy between $V_{\rm max}$ and the inclination $i$ is broken,
thanks to our 3D method, but the seeing leaves a significant correlation between $V_{\rm max}$  and the turnover radius $r_{\rm t}$.
The presence of this degenerancy is data specific and seeing specific, not a generic feature of the algorithm. \label{fig:correl}}
\end{figure*}

Figure~\ref{fig:correl} shows the joint distributions for the radius, PA, inclination, maximum velocity, and dispersion parameters.
 The estimated parameters and their respective $1~\sigma$ error are shown as
a solid line and dashed line, respectively.
This figure shows a clear covariance between the turnover radius and the asymptotic velocity $V_{\rm max}$, and  a small covariance between the inclination and $V_{\rm max}$.  The users of \galpak\ are strongly advised to   confirm the convergence of the parameters using diagnostics similar to Figure~\ref{fig:mcmc}
and   to investigate possible covariance in the parameters, as these tend to be data specific,
using diagnostics similar to Figure~\ref{fig:correl}.


\section{Tests with mock data cubes}
\label{section:simu}

\begin{table}
\caption{Range of parameters for the \Nideal\ mock galaxies  \label{table:grid}}
\centering
\begin{tabular}{lrlr}	
Parameter & Grid Values &\\
\hline
Flux ($10^{-17}$\flux) & 3, 6, 10, 30 \\
Seeing (") & 0.6, 0.8, 1.0, 1.2 \\ 
Redshift & 0.6, 0.9, 1.2 \\ 
$R_{1/2}$ (kpc) & \sizes  \footnote{Exact value  to satisfy the size-velocity scaling relation \citep{DuttonA_11a}. } \\
$R_{1/2}$ (") & \sizesArc \footnote{Exact value will depend on the redshift.} \\
Incl. (deg) &  20, 40, 60, 80   \\ 
PA (deg) & 130 \\
$r_{\rm t}$ (") & 0.1--0.3\footnote{Exact value to satisfy the scaling relation between the galaxy size and the inner gradient \citep{AmoriscoN_10a} using $r_{\rm t}=R_{\rm d}/1.8$.}  \\
$V_{\rm max}$  (\kms) & 110, 200,  280   \\
$\sigma_o$ (\kms) & 20, 50, 80 \\ 
\hline
\end{tabular}
\end{table}

In order to characterize the performances and limitations of the GalPaK$^{3D}$ algorithm statistically, we generated a set of  \Nideal\ cubes again with a MUSE configuration over a grid of parameters listed in Table~\ref{table:grid}. The synthetic cubes were 
generated  with noise typical to a 1~hr exposure with MUSE corresponding to a pixel noise of $\sigma=5\times10^{20}$~\flux~\AA$^{-1}$.
We use a range of inclinations $i$ from 20$^\circ$ to 80$^\circ$. 
We use a range of disk sizes, with half-light radii $R_{1/2}~=$~\sizesArc\  corresponding to a $R_{1/2}$ of \sizes~kpc, covering the range of observed sizes at $z\sim1$ \citep[e.g.][]{TrujilloI_06a,WilliamsR_10a,DuttonA_11a}. 

For each of the galaxy sizes, we use the $V_{\rm max}$-$R_{1/2}$ scaling relation \citep[Equation 8 of][]{DuttonA_11a}
and its redshift evolution \citep[Equation 5 of][]{DuttonA_11a} to set the rotation kinematics ($V_{\rm max}$).
In particular, the  sizes $R_{1/2}=$\sizes~kpc correspond to $V_{\rm max}$ values ranging from $\sim100$ to 250~\kms.
We use  ``arctan'' rotation curves to generate our mock data cubes, and we have verified that our results remain the same with ``exponential'' rotation curves.

We use the scaling relation between the turnover radius $r_t$ and the disk scale length $R_{\rm d}$ that exists 
for disk galaxies \citep[e.g. Figure~1 of][]{AmoriscoN_10a} to set the turnover radius $r_{\rm t}$. 
In particular, we set $r_{\rm t}$ to $R_{\rm d}/1.8$ where the 1.8 factor~\footnote{For `exponential' rotation curves, one should set $r_{\rm t}$ to $R_{\rm d}\times0.9$ in order to satisfy the scaling relation; for `tanh' rotation curves, one should set  $r_{\rm t}$ to $R_{\rm d}\times1.25$.} is determined empirically for the arctan rotation curve
to satisfy the linear correlation between the galaxy disk scale-length $R_{\rm d}=R_{1/2}/1.68$  and $R_{\Omega}$, defined as the radius $r$ where $V(r)=2/3\;V_{\rm max}$ \citep{AmoriscoN_10a}.

For each of the galaxy sizes, the  disk thickness is  $h_z=0.15\,R_{1/2}$, i.e. ranging from 0.4 to 1.3~kpc,
bracketing the average values of $h_z\sim1$~kpc, found for high-redshift edge-on/chain galaxies \citep{ElmegreenB_06a}.

We used fluxes for an [OII] ($\lambda$3727) emission line, expected to lie in the MUSE spectral range at redshifts between 0.6 and 1.2,
with integrated fluxes from $3\times10^{-17}$~\flux\ to $3\times10^{-16}$~\flux\ corresponding to the range of observed values  
\citep[e.g.][and references therein]{BaconR_15a,ComparatJ_15a}.
We use a  constant noise value per pixel of $\sigma=5\times10^{-20}$~\flux~\AA$^{-1}$, in order to simulate the noise level of a 1~hr exposure,
but we stress that the algorithm accepts variance/noise cubes to account for pixel-to-pixel noise variations.
In addition, we generated  cubes  with very high S/N (S/N=100, flux$=3\times10^{-15}$~\flux) and with a seeing typical of AO conditions,  with a PSF FWHM of 0\farcs15.  These will serve as reference data sets.


\subsection{Surface Brightness and Signal-to-noise Ratio}

One could imagine that the S/N in the recovered parameters be a function of the average  S/N pixel$^{-1}$, or the apparent SB since the observed central SB  scales directly with the SNR in the central pixel. 
But clearly the compactness of the object with respect to the seeing plays a large role \citep[as discussed in][]{DriverS_05a,EpinatB_10a}.
Very  compact objects (compared to the beam or the PSF)  have high SB by definition (and high S/N pixel$^{-1}$), 
but the morphology and/or kinematic
information may be lost owing to the beam smearing.    On the other hand,
very extended objects have low surface brightness (and low S/N pixel$^{-1}$), but have many pixels in the outer regions (with low S/N),
where most of the information on the galaxy is located and not affected by the beam.

Before illustrating this point, it is important to define commonly used terms such as the SB of galaxies.
From any light profile $I(r)$ such as given by Equation~\ref{eq:Ir}, there are many ways to define galaxy SB, such as
$I_{\rm e}$ the SB   at the effective radius $R_{\rm e}$,
$I_{\rm o}$ the intrinsic  SB at the central pixel, 
 $A_{\rm o}$ the observed  SB at the central pixel, and
SB$_{1/2}$ the average SB within  the intrinsic half-light radius $R_{1/2}$:
\begin{equation}
{\rm SB}_{1/2,\rm conv}\equiv\frac{0.5\,F_{\rm tot}}{\pi R^2_{1/2,\rm conv}}.\label{eq:SBconv}
\end{equation}
where $F_{\rm tot}$ is the galaxy total flux.
A related quantity to Equation~\ref{eq:SBconv} is the observed SB, defined as :
\begin{equation}
{\rm  SB}_{1/2,\rm obs}\equiv \frac{0.5\, F_{\rm tot}}{S_{1/2,\rm obs}}\label{eq:SBobs}
\end{equation}
where $F_{\rm tot}$ is the galaxy total flux
and $S_{1/2,\rm obs}$ the galaxy apparent area given by
 $\equiv \pi a\,b$ where $a$ and $b$ are the observed major and minor semiaxes, respectively, of the galaxy.
The relations between these various definitions are described in  the appendix.

\begin{figure*}
\centering
\subfigure[]{
\includegraphics[width=8cm,height=8cm]{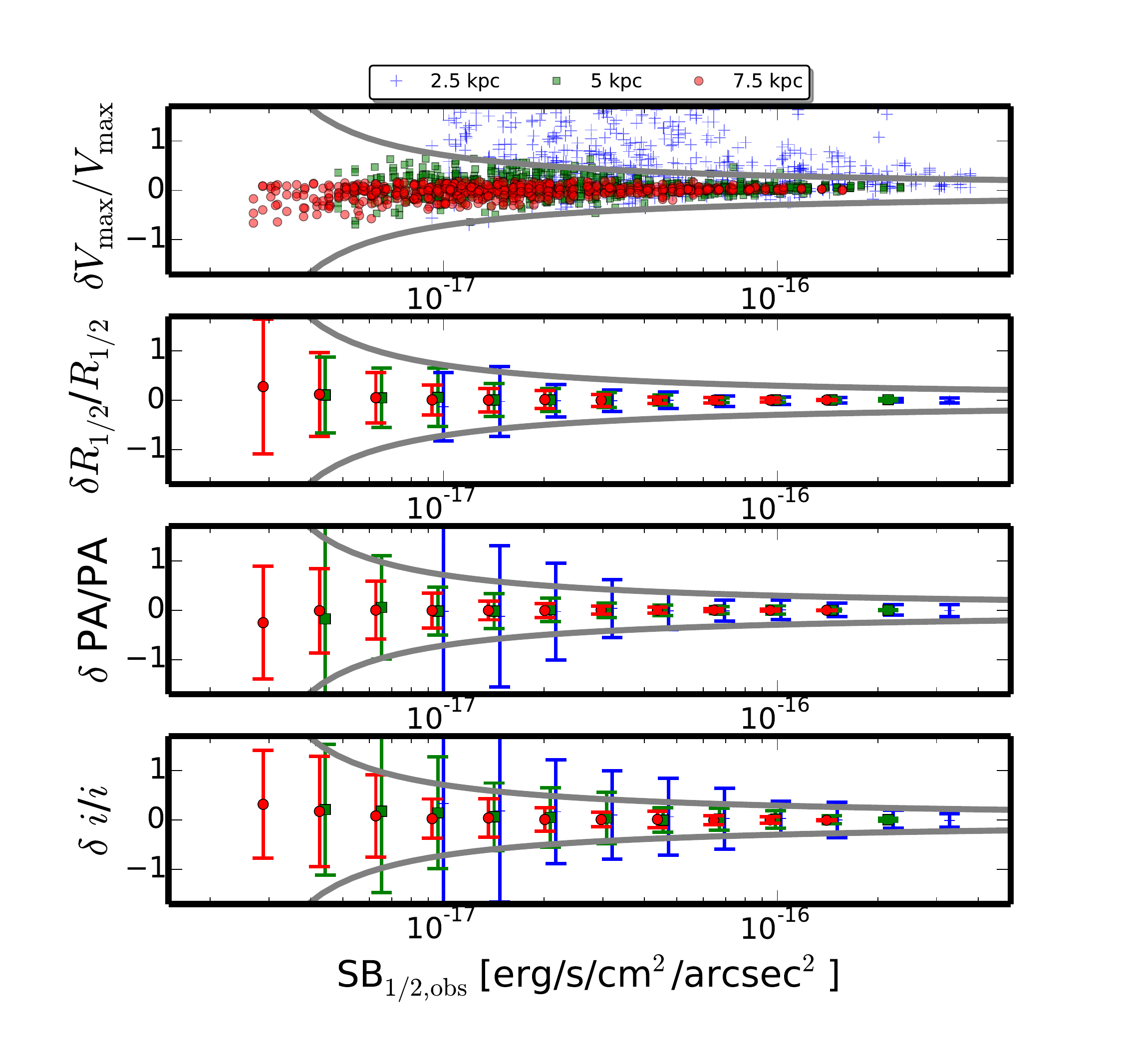}
}
\subfigure[]{
\includegraphics[width=8cm,height=8cm]{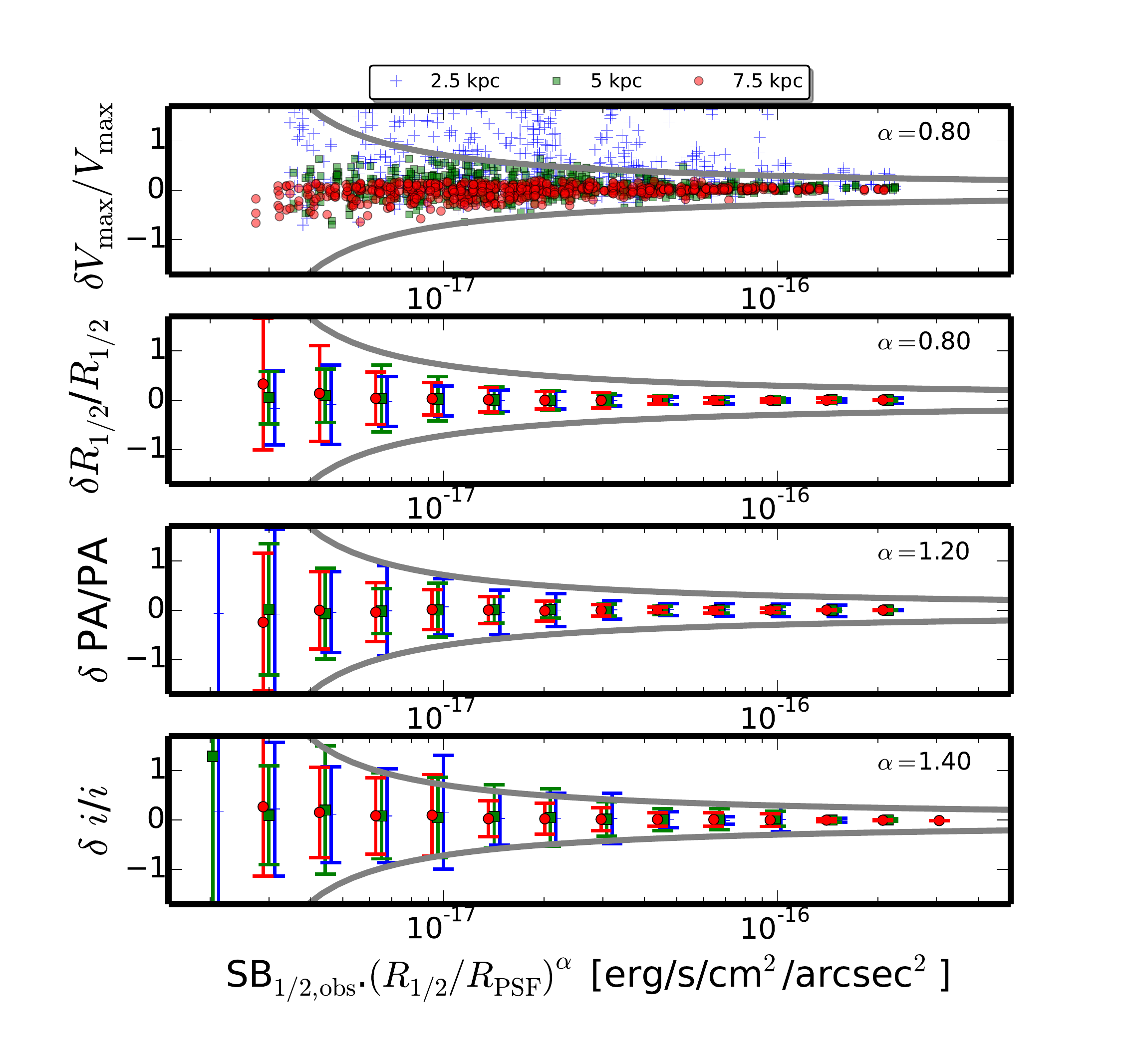}
}
  \caption{ Relative errors on the estimated parameters $\delta p/p$, defined as 
 $(p_{\rm fit}-p_{\rm in})/p_{\rm in}$. 
 Each row  shows $\delta p/p$ for the maximum circular velocity $V_{\rm max}$, the size $R_{1/2}$, 
the PA,  and inclindation $i$ from top to bottom, respectively.
The crosses, squares, and circles represent the three subsamples with sizes $\sim$ \sizes\ kpc, respectively (Table~\ref{table:grid}).
The relative errors in for the morphological parameters (size, PA, inclination) are binned.
{\it Left}(a): Relative errors as a function of central surface brightness SB$_{1/2,\rm obs}$ in erg/s/cm$^2$/arcsec$^2$.
{\it Right}(b): Relative error as a function of central SB, SB$_{1/2,\rm obs}$, times
$(R_{1/2}/R_{\rm PSF})^{\alpha}$, where $R_{1/2}$ is  the galaxy
{ intrinsic} half-light radius  and $R_{\rm PSF}$  the PSF half-light radius.
We found, empirically (see text), that $\alpha$ is approximately 0.8, 1.2, and 1.4 for the size, PA, and inclination parameter, respectively.
These values are close to the expectation of $-1.0$ of  Equation~\ref{eq:refregier} (gray lines) derived for morphological parameters in imaging data by \citet{RefregierA_12a}.
The relative error in $V_{\rm max}$ does not follow the expected relation  and is subject to strong systematics for the smallest and largest mock galaxies (crosses). This is due to $V_{\rm max}$ being constrained in the outer parts
of the galaxy, where the S/N is thus not sufficient for the compact galaxies or where the mock cube is too small for the largest galaxies.  
  \label{fig:SBplot}
}
\end{figure*}

To illustrate the point made at the beginning of this section, we show in Figure~\ref{fig:SBplot}(a)
the relative errors $\delta p/p\equiv(p_{\rm fit}-p_{\rm in})/p_{\rm in}$ on some of the estimated parameters 
 for our mock data cubes generated in Section~\ref{section:simu}  
as a function of central SB, SB$_{1/2,\rm obs}$ (defined in Equation~\ref{eq:SBobs}).
Each row shows the relative errors  for   the maximum circular velocity $V_{\rm max}$,
the size $R_{1/2}$,  the PA, and inclination $i$ from top to bottom, respectively.
The crosses, squares and circles represent the three subsamples with sizes $\sim$ \sizes\ kpc, respectively.
One sees that the errors in the morphological parameters (size, PA, inclination) do increase toward low SBs, but
the threshold point at which the relative errors reach $\sim$100\%\ depends on the galaxy size, represented by the symbols.
This illustrates the well-known fact that very extended objects have low surface brightness (and low S/N pixel$^{-1}$) but have many pixels in the outer regions that contain useful information.

As argued at the beginning of this section  and demonstrated in Figure~\ref{fig:SBplot},
 SB alone might not be sufficient to determine the S/N in the fitted parameters, but the
compactness of the galaxy with respect to the beam also plays  an important role.
 In Figure~\ref{fig:SBplot}(b), we show   the relative error $\delta p/p$    with respect to the observed SB$_{1/2, \rm obs}$
 times the size-to-PSF ratio $(R_{1/2}/R_{\rm PSF})^{\alpha}$.
 The symbols correspond  to galaxy subsamples with various sizes as in Figure~\ref{fig:SBplot}(a).  
The index $\alpha$ was found to be empirically $\sim1$ in order to have the relative errors
for each of the subsamples follow a similar trend and may differ sightly for each of the parameters $p$. In fact, we
find that $\alpha$ is  approximately 0.8, 1.2, and 1.4 for the size, PA, and inclination parameter, respectively.

These empirical results can be explained by the following arguments. 
The apparent  SB within the half-light radius   SB$_{1/2,\rm conv}$ (Equation~\ref{eq:SBconv})
and the observed SB$_{1/2, \rm obs}$ (Equation~\ref{eq:SBobs}) are proportional to the SB (or S/N) of the central pixel, $A_{\rm o}$,
as shown in the Appendix (Equation~\ref{eq:appendix}).
In the case of no PSF convolution, \citet{RefregierA_12a} showed that (their Equation 12)
 the relative error $\sigma(a)/a$ on morphological parameters (its major-axis $a$) scales inversely to the central $I_{\rm o}$ where
 $I_{\rm o}$ is the intrinsic central SB  (Equation~\ref{eq:Io:exp}--\ref{eq:Io:gau}).
In the presence of a PSF convolution,    Equation~16 of \citet{RefregierA_12a} ---which applies here--- 
shows that the relative errors on the major-axis $a$ scale as
\begin{eqnarray}
\frac{\sigma(a)}{a}&\propto& A_{\rm o}^{-1}(1+R_{\rm PSF}^2/R_{1/2}^2).\label{eq:errorp}
\end{eqnarray}
where   $R_{\rm PSF}$ is  the radius of the PSF ($R_{\rm PSF}\equiv$ FWHM/2) and $R_{1/2}$ the {intrinsic} half-light radius.

In our cases, for high-redshift galaxies,  the ratio $R_{\rm PSF}/R_{1/2}$ is $\simeq1.0$ and after performing a Taylor
expansion around
$R_{\rm PSF}/R_{1/2} \sim (1-x)$ with $x\equiv (R_{1/2}-R_{\rm PSF})/R_{1/2}$ and $|x|<<1$,
one finds that   the factor $(1+R_{\rm PSF}^2/R_{1/2}^2)$
is approximately $\sim 2\;(1-x)\sim2 \;R_{\rm PSF}/R_{1/2}$.
Hence, Equation~\ref{eq:errorp}   on the errors in the major-axis $a$ becomes
in the regime where $R_{\rm PSF}/R_{1/2}\simeq1.0$:
\begin{eqnarray}
\frac{\sigma(a)}{a}&\propto&  \left(\frac{R_{1/2}}{R_{\rm PSF}}A_{\rm o}\right)^{-1}\nn\\
& \propto& \left(\frac{R_{1/2}}{R_{\rm PSF}} {\rm SB}_{1/2,\rm obs} \right)^{-1}\label{eq:refregier},
\end{eqnarray}
which shows that the quality of the estimated morphological parameters will depend on both the pixel S/N (or SB) and 
  the galaxy compactness with respect to the beam, ${R_{1/2}}{R_{\rm PSF}}$, as shown in Figure~\ref{fig:SBplot}(b) 

In both Figure~\ref{fig:SBplot}(a) and~\ref{fig:SBplot}(b), the gray solid lines show the expected behavior for the morphological parameters (Equation~\ref{eq:refregier})  and one sees that they agree  better with the  mock data in the right panels for the morphological parameters.
This shows that the \citet{RefregierA_12a} formalism describes the relative errors on the morphological parameters
(size, PA, and inclination) relatively well, as a first approximation. 
We note that Equation~\ref{eq:refregier} is only an approximation to Equation~\ref{eq:errorp} when $R_{1/2}/R_{\rm PSF}\simeq1$ and  that
there might be other dependencies for the other morphological parameters, namely, for the PA and for the inclination. 
Here  we refer the reader to Table~1 of \citet{RefregierA_12a} and their Appendix for further details;
 it is beyond the scope of this paper to present a full 3D derivation
of the \citet{RefregierA_12a} formalism.

Contrary to the morphological parameters, the errors in the kinematic parameter $V_{\rm max}$
show strong positive (negative) biases in the smallest (largest) mock galaxies, represented by the crosses (circles)
respectively in the top panel of Figure~\ref{fig:SBplot}(b).
The positive bias for for the most compact galaxies (crosses) with respect to the beam
can be understood because the $V_{\rm max}$  information is located mostly in the outer parts
of the galaxy, where the S/N is too low.
The negative bias for the largest galaxies (1" in $R_{1/2}$) at low SB is likely due to the spatial cut of our mock cubes  being too small.

We will return to the reliability of $V_{\rm max}$ in section~\ref{section:vmax} and now turn to
a more detailed discussion on the reliability of the parameters (size, inclination, disk velocity dispersion, and $V_{\rm max}$).
While we used an arctan rotation curve, we note that the following results were found to be identical when we used an `exponential' rotation curve.

\subsection{Reliability of morphological parameters}

We have shown in the previous section with Figure~\ref{fig:SBplot} that the relative errors on the half-light radius follow
appoximately the expectation from the  \citet{RefregierA_12a}  formalism. 
Here  we investigate whether the relative errors depend on some of the other parameters, such as inclination,
seeing, and size.

\begin{figure*}
\centering
\includegraphics[width=19cm]{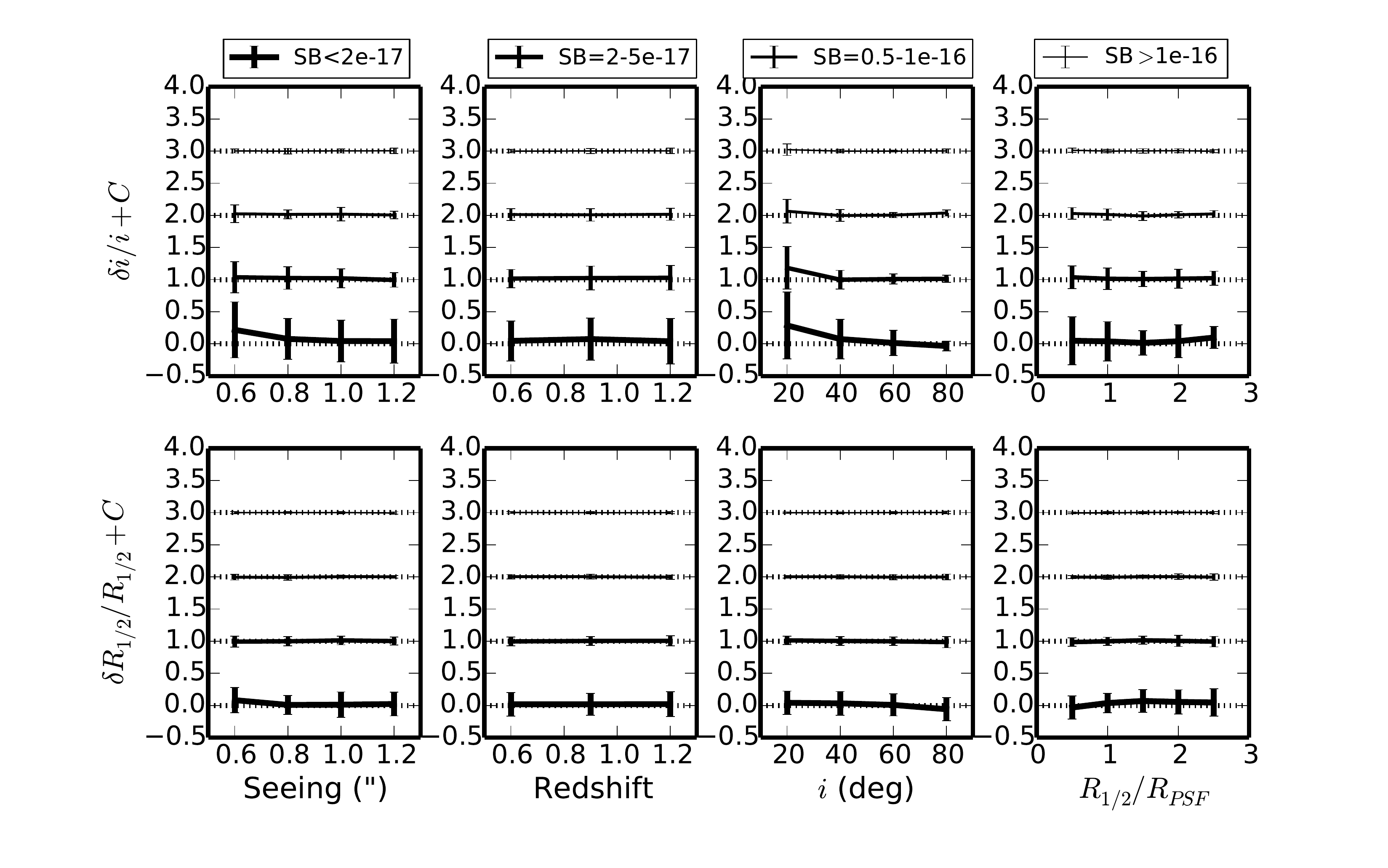}
\caption{Relative error  $\delta p/p$ ---defined as 
 $(p_{\rm fit}-p_{\rm in})/p_{\rm in}$---  for the parameters $R_{1/2}$ (top) and inclination $i$ (bottom) 
as a function of seeing, redshift, inclination $i$, and size-to-psf ratio $R_{1/2}/R_{\rm PSF}$ (from left to right).
 The  curves   with increasing thickness  correspond to subsamples with different SB levels
from SB $<2\times10^{-17}$,  $2\times10^{-17}<$ SB $<5\times10^{-17}$,  $5\times10^{-17}<$ SB $<1\times10^{-16}$,  SB $>1\times10^{-16}$
\flux~arcsec$^{-2}$  respectively, where the zero point (dotted line) has been offset  for clarity purposes.
  SB is the surface brightness within the observed half-light radius SB$_{1/2,\rm obs}$ times the seeing-to-size ratio $R_{1/2}/R_{\rm PSF}$, as in Figure~\ref{fig:SBplot}.
The data points represent the median,  and the size of the error bars represent the standard deviation for each of  the subsample.
One sees that the \galpak\ algorithm recovers the morphological parameters irrespectively of seeing, redshift, and/or intrinsic size.
\label{fig:matrix_size}}
\end{figure*}

Figure~\ref{fig:matrix_size} shows the relative errors  $(p_{\rm fit}-p_{\rm in})/p_{\rm in}$ for several key parameters $p$.
The bottom (top) row shows the result for the size parameters $R_{1/2}$ (inclination $i$), respectively,
as a function of seeing, redshift, inclination, and size-to-psf ratio $R_{1/2}/R_{\rm PSF}$.
The black curves  with increasing thickness  correspond to subsamples with different SB levels (labeled) 
where the zero point (dotted line) has been offset for clarity purposes.
The data points represent the median,  and the size of the error bars represent the standard deviation for each of  the subsamples, 
where we have typically $\sim 100$ mock cubes per bin.
We note that the median standard deviations on the parameters (from the posterior distributions) tend to be  
within  20\%\ of these binned standard deviations.

From this figure, one sees that the \galpak\ algorithm recovers the intrinsic half-light radius $R_{1/2}$ irrespectively of seeing, redshift, and/or intrinsic size. Note that the relative errors with respect to size-to-seeing ratio at a fixed SB follow roughly the expectation from Equation~\ref{eq:errorp},
where the factor  $1+(R_{\rm PSF}/R_{1/2})^2$  saturates to unity in our regime with $R_{1/2}/R_{\rm PSF}$ $\sim1$ to 2.5.
These  results are not affected by the choice of the SB profile (Sersic $n$).\footnote{A  curve-of-growth analysis on the two-dimensional flux map can sometimes yield a constraint on the Sersic index $n$ and a more accurate determination of the intrinsic half-light radius ($R_{1/2}$).}

From the top row in Figure~\ref{fig:matrix_size}, one  sees that the input inclination is recovered
except at the two smallest fluxes and for the more face-on cases. 
The reason that the algorithm can recover the inclination well is that the algorithm breaks  the traditional degeneracy between $V_{\rm max}$ and $i$ using the SB profile (i.e. the axis ratio $b/a$) whereas traditional
methods fitting the kinematics on velocity fields have a strong degeneracy between $V_{\rm max}$ and the inclination $i$.

\begin{figure*}
\centering
\includegraphics[width=19cm]{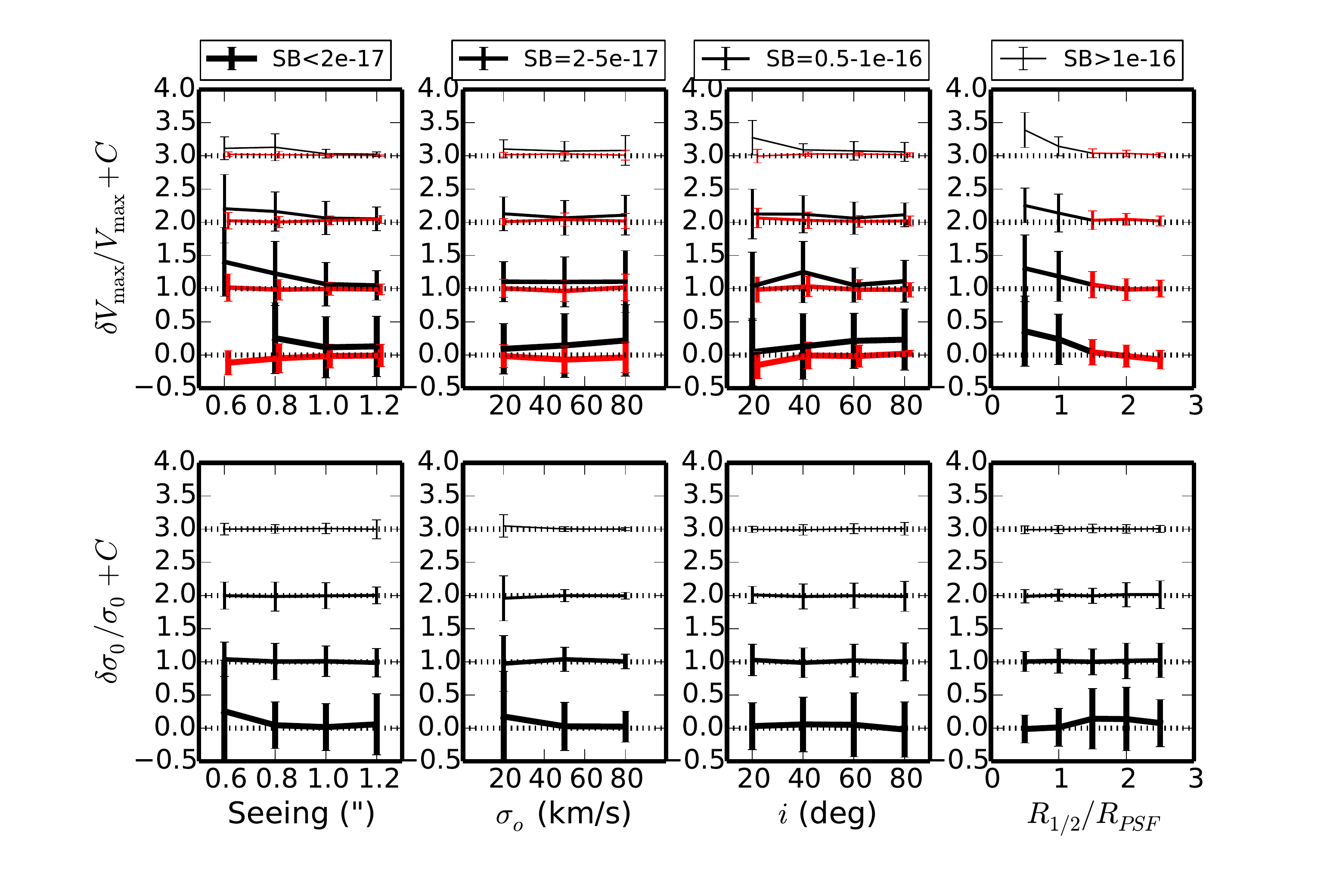}
\caption{Relative error  $\delta p/p$ ---defined as 
 $(p_{\rm fit}-p_{\rm in})/p_{\rm in}$---  for the kinematic parameters $V_{\rm max}$ (top) and $\sigma_o$ (bottom) 
as a function of seeing, disk dispersion $\sigma_o$, inclination $i$, and size-to-PSF $R_{1/2}/R_{\rm PSF}$ (from left to right).
The curves as a function of redshift are not shown because the relative errors  do not depend on this parameter as in Figure~\ref{fig:matrix_size}.  
For the $V_{\rm max}$ parameter, the black (red) curves show the results when $R_{1/2}/R_{\rm PSF}$ is less (greater) than \rmin, respectively.
One sees that the \galpak\   algorithm recovers the kinematic parameters irrespectively of seeing and redshift, provided that the galaxy is not too compact with $R_{1/2}/R_{\rm PSF}$  larger than \rmin.
\label{fig:matrix_Vmax}}
\end{figure*}

\subsection{Reliability of  kinematic parameters}
\label{section:vmax}

Figure~\ref{fig:matrix_Vmax} shows the relative errors  $\delta p/p\equiv(p_{\rm fit}-p_{\rm in})/p_{\rm in}$ for 
  the parameters $V_{\rm max}$ (top row) and disk dispersion $\sigma_o$ (bottom row)
as a function of seeing, $\sigma_o$, inclination, and size-to-PSF ratio $R_{1/2}/R_{\rm PSF}$.
The curves as a function of redshift are not shown, because the relative errors  do not depend on this parameter as in Figure~\ref{fig:matrix_size}. 
The black curves  with increasing thickness  correspond to subsamples with different SB levels (labeled)  
 where the zero point (dotted line) has been offset  for clarity purposes.
The data points represent the median,  and the size of the error bars represents the standard deviation for each of  the subsamples, as in Figure~\ref{fig:matrix_size}.

 Figure~\ref{fig:matrix_Vmax}(top) shows that the \galpak\ algorithm  recovers the maximum velocity $V_{\rm max}$
 irrespectively of seeing, disk dispersion, and redshift (now shown)
provided that the galaxy is not too compact.  For small galaxies with $R_{1/2}/R_{\rm PSF}$ less than \rmin,
the figure shows that it is increasingly difficult to estimate the correct values for the most compact galaxies,
with large uncertainties and significant overestimations of this parameter.
This result was already pointed out in \citet[][their Figure 13]{EpinatB_10a} using 2D kinematic models. \citet{EpinatB_10a} also noted that using a simple flat rotation curve to model the disk, the maximum velocity $V_{\rm max}$ can be recovered with an accuracy better than 25\%, even when $R_{1/2}/R_{\rm PSF}$ is less than about $\sim2$.

Figure~\ref{fig:matrix_Vmax}(bottom) shows  the \galpak\ algorithm recovers the disk dispersion irrespectively of seeing
  and redshift (not shown).
Given the instrumental resolution of MUSE used here ($R\simeq130$~km/s), small dispersions are more difficult to recover. 
We note that the local dispersion is rather sensitive to the instrument LSF FWHM, as one might expect.
The user can specify more than one type of LSF (Gaussian or Moffat), and a user-provided
vector can be specified if the parametric LSF is not sufficient to describe the instrument LSF.

\subsection{A note regarding the PSF accuracy}
\label{section:PSF}

One could argue that our results are driven by the fact that we use the exact same PSF (in 3D) as the one used to generate these modeled galaxies.
To test the reliability of the algorithm in more realistic situations, when the PSF FWHM is not known accurately,
we ran the algorithm on the same set of data cubes with a random component added to the 
FWHM of the PSF given by a normal distribution with $\sigma=0.1$, corresponding to uncertainties in the FWHM of $\sim20$\%.
We found that the accuracy of the spatial kernel (PSF) has little impact on the recovered parameters.
On the other hand, we find that the shape of the PSF is more critical especially for the morphological parameter such as the axis ratio $b/a$ (or the inclination). We note that sophisticated tools exist to determine the PSF from faint stars in data cubes such as the algorithm of \citet{VilleneuveE_11a}.

To conclude this section, our algorithm is able to recover the morphological and kinematic parameters from synthetic data  cubes over a wide range of seeing conditions provided that the galaxy is not too compact and has a sufficiently high SB.
Thus,  for galaxies to be observed with MUSE in the wide-field mode in 1~hr exposure and no AO, 
we find that the algorithm should perform well provided that the SB is greater
 than a few $\times10^{-17}$~\flux~arcsec$^{-2}$ and  
as long as the the size-to-seeing ratio $R_{1/2}/R_{\rm PSF}$ is larger than 1.5
(or  $R_{1/2}/$FWHM $>0.75$).


\section{Application on Hydrodynamical Simulations}
\label{section:Leo}

In the previous section we validated the algorithm on synthetic or mock data, 
which have by definition no defects, i.e. are perfectly regular and symmetric.
In order to validate the algorithm on more realistic data, 
we now analyze the performance of the algorithm on data cubes created from simulated galaxies generated  
from a hydrodynamical simulation (Michel-Dansac et al., in prep.).
This   is intended to validate the algorithm in the presence of systematic deviations from the disk model.

\subsection{From Hydrodynamical Simulations to Data cCubes}

The simulation used in this work comes from a set of cosmological zoom
simulations, each targeting the evolution until redshift 1 of a single
halo and its large-scale environment. The full sample of simulations
is presented in details in Michel-Dansac et al., in prep. Here we focus on
one output of one simulation to complement the  test cases from section~\ref{section:simu}
with a more realistic, intermediate redshift, star-forming disk galaxy.  

The simulations have been run with the Adaptative Mesh Refinement
code \texttt{RAMSES} \citep{TeyssierR_02a} using the standard zoom-in
resimulation technique to model a disk galaxy in a cosmological context.
Each simulation has periodic boundaries and nested levels of
refinement in a zoom region around the targeted halo, in both DM and
gas. The refinement strategy is based on the quasi-Lagrangian
approach. The simulation zooms in a dark matter halo inside a $20~h^{-1}$~Mpc
comoving box, achieving a maximum resolution of $\sim 200$~pc. 
The virial mass of the dark matter halo is approximately $3\times 10^{11}
\msun$ at $z=1$, sampled with roughly $600,000$ particles. 

The simulation implements standard prescriptions for various physical
processes crucial for galaxy formation: star formation, metal
enrichment, and kinetic feedback due to Type II supernovae 
\citep{DuboisY_08a}; 
metal advection, metallicity- and density-dependent cooling;
 and UV heating due to cosmological ionizing
background  \citep[see][for more details on similar simulations
but focusing on $z=0$ Milky-Way-type galaxies]{FewC_12a}.

\begin{figure*}[t]
 \centering
 \includegraphics[width=18cm]{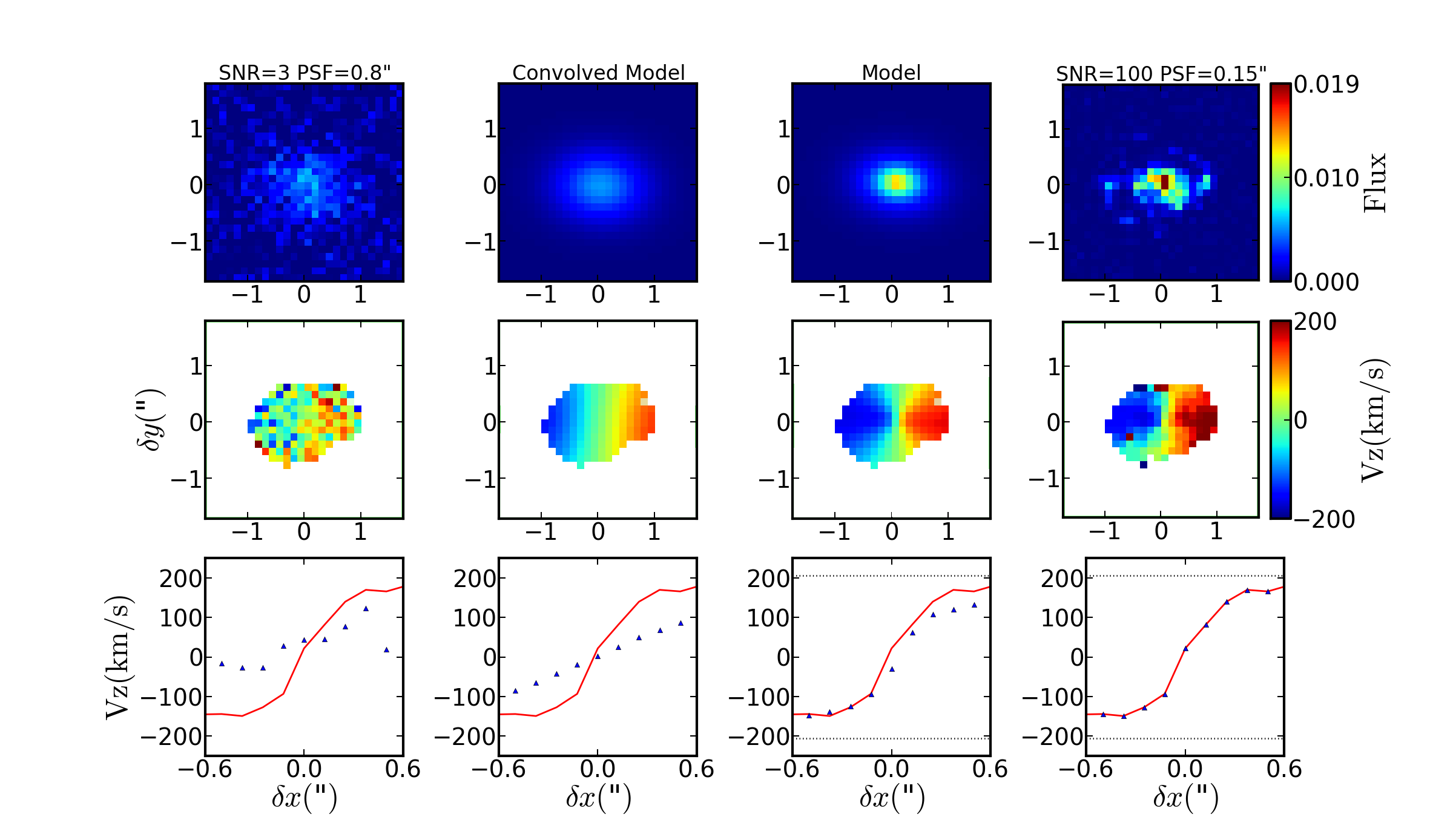}
 \caption{Application of the MCMC algorithm  on a disk galaxy generated with the AMR code \texttt{RAMSES} \citep{TeyssierR_02a}
and `observed' with  a seeing of 0\farcs8 (FWHM). 
The maximum SNR is $\sim$3 in the brightest pixel. 
The top, middle, and bottom rows show the  flux map, the velocity map  and the apparent velocity profile $V_z(r)$ across the major axis, respectively.
From left to right, the panel columns show the data, the convolved model, the modeled disk (free from the PSF),
 and the high-S/N high-resolution reference data (PSF=0\farcs15 and S/N=100). 
 In the bottom panels, the solid red curves  correspond to the reference case, the triangles represent the apparent rotation curve, and  the dotted lines show the apparent maximum line-of-sight velocity  $V_{\rm max}\,\sin i$.
One sees that the the velocity profile from the modeled disk (third column) is in good agreement with the reference data (solid line) at 0\farcs15 resolution.}\label{fig:Leo}
\end{figure*}

The simulated galaxy is a typical $z=1$ star-forming galaxy with
$M_{\star} = 3\times 10^{10}\msun$ and a gas fraction of 0.33. The
galaxy exhibits a disk morphology with spiral arms as seen in Figure~\ref{fig:Leo} (top right panel).

From the output of the hydro-simulation, we generated a data cube
with the  Spectrograph for INtegral Field Observations in the Near
Infrared (SINFONI) instrumental resolution and pixel size (0\farcs125 pixel$^{-1}$ and 2 \AA~pixel$^{-1}$)
using  the star formation rate (SFR) and metallicity information in each cell. 
To construct the mock data cube, the simulated galaxy is artificially
placed at $z=1.3$ ($\lambda_c \simeq 1.5 \mu $m for \Ha) and rotated with an inclination of $60^{\circ}$.
Star-forming cells are selected by computing the mass of young stars
inside each cell of the galaxy. Then, we convert this star formation
rate into \Ha\ flux using the \citet{KennicuttR_98a} calibration.
For each cell, we also compute the flux in the \NII\ line
from the values of the \Ha\ flux and the oxygen abundance  
following the calibration given by \citet{ContiniT_09a}.
For each spatial element or spaxel,  we sum the contribution (to the spectrum) of each cell along the line-of-sight.
Each contribution has its own line of sight velocity, which blueshifts or redshifts the lines. 
The  line width in the spectrum is then due to the sum or integral over the cells,
which is then convolved with the instrumental profile.

We generated seeing-convolved cubes with seeing of  0\farcs50, 0\farcs65,  0\farcs80, 1\farcs0 and 1\farcs2 (corresponding to typical values in the NIR with SINFONI)
and 0\farcs15 (corresponding to adaptive optic assisted observations)
 and added noise corresponding to a given max S/N pixel$^{-1}$.
  Cubes generated with a SNR equal to 100 and a seeing of 0\farcs15 are used as reference cubes.
The final cube size is $28\times28\times30$ (in $x$, $y$, $\lambda$ directions), but we also produce another set of cubes of size $28\times28\times200$ pixels to allow sufficient wavelength baseline for our custom  line-fitting algorithm that was used to produce the 2D velocity maps shown in Figure~\ref{fig:Leo}.

\subsection{Application of the algorithm}

Figure~\ref{fig:Leo} shows the results of the GalPaK$^{3D}$ algorithm for a seeing of 0\farcs8 and a minimum SNR pixel$^{-1}$ of 3 in the brightest pixel.
As in Figure~\ref{fig:SNR3}, 
the top, middle, and bottom rows show the  flux map, the velocity map  and the apparent velocity profile $V_z(r)$ across the major axis, respectively.
From left to right, the panel columns show the data, the convolved model, the modeled disk (free from the PSF),
 and the high-SNR high-resolution reference data (PSF=0\farcs15 and SNR=100).  
In the bottom panels, the solid red curves  correspond to the reference rotation curve (obtained from the reference data set), 
and the triangles represent the apparent rotation curve.
By comparing the two, one sees that the algorithm is able to recover the kinematics (third column) in a regime where traditional 2D methods (left most column) tend to be noisier. In other words, 
the recovered kinematics from the modeled disk (intrinsic or unconvolved model) shown in the third column
 is in good agreement with the reference data (last column) in spite of the lower spatial resolution (0\farcs8) and the lower S/N in the data set.

We ran the \galpak\ algorithm on the data cubes, setting the  rotation curve $v(r)$ to an`` arctan'' profile and setting the Sersic index $n$ to $1.0$.\footnote{We also
ran the algorithm  with ``gaussian'' profiles with $n=0.5$ leading to very similar results.}
From the cube with a S/N of 100, the inclination found by the \galpak\ algorithm is $58^\circ\pm2^\circ$, and 
the half-light radius $R_{1/2}$ is $\sim 3.4\pm0.1$~kpc (or $\sim$ 0\farcs4), and its asymptotic maximum velocity $V_{\rm max}$  is $\sim 215\pm10$~\kms,
placing it close to the $z\sim1.5$ size-velocity relation of \citet{DuttonA_11a}.
The asymptotic maximum velocity is close to the one extracted directly from the simulation, which is 235 \kms.

\begin{figure}
 \centering
 \includegraphics[width=8cm]{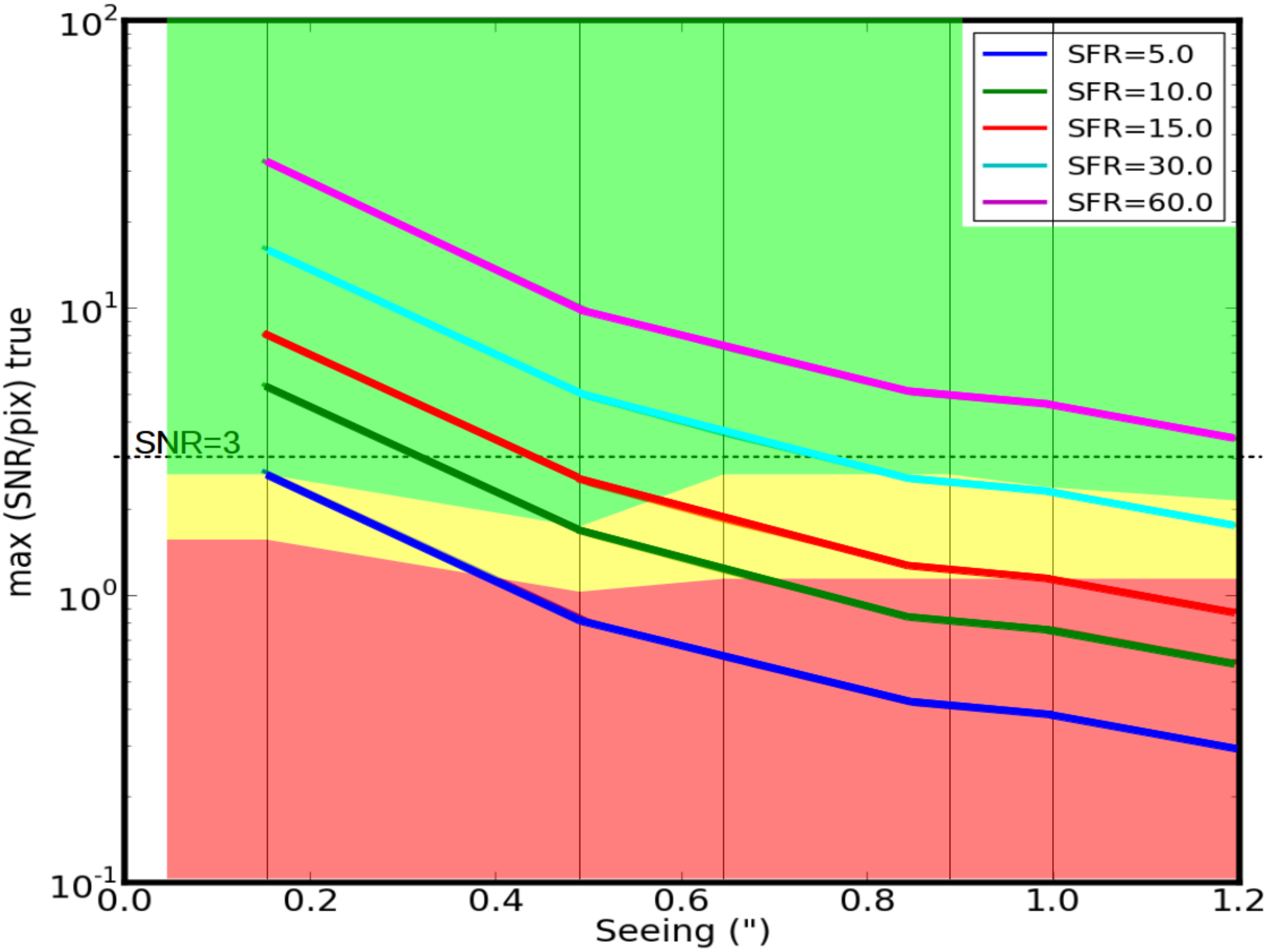}
 \caption{S/N pixel$^{-1}$ as a function of seeing (FWHM) for our SINFONI data cube simulated for a 2~hr exposure time. 
The lines correspond to a given SFR at $z=1.3$ from SFR=5 to 60 \mpy. 
The simulated cubes (generated from the hydrodynamical simulation)
 can be reasonably well fitted by our algorithm provided that the S/N pixel$^{-1}$ (at the central region)
 is greater than 3, irrespectively of seeing.
This diagram applies to galaxies with inclinations around $\sim60^\circ$.  \label{fig:global}}
\end{figure}

We repeated the exercise on this simulated galaxy varying the luminosity (SFR in our case) where the noise level is set for a given exposure time corresponding to a 2 hr integration with the SINFONI instrument.  Figure~\ref{fig:global} shows the maximum signal to noise per pixel (solid lines)
as a function of the seeing FHWM for five fixed SFRs, 5, 10, 15, 30, and 60 \mpy, respectively. 
The green region shows the parameter space where the algorithm is able to recover the kinematics parameters within 20\%,
from the value determined in the high-S/N cube.
The yellow region shows the parameter space where the algorithm is marginally able to recover the kinematics parameters,
i.e. within 20\%--40\%
The red region shows the parameter space where the algorithm is unable to recover the kinematics parameter, where the
relative error is larger than 40\%.
This plot shows that the kinematic parameters can be well estimated irrespectively of seeing, provided that the SNR is above a critical value  (3 in this case).
Consequently, when the PSF FWHM is slightly below the original scientific goal, the optimal observing strategy is to integrate longer. 

In the background-limited regime, the S/N per pixel scales as $\propto  \sqrt{t_{\rm exp}}$, where $t_{\rm exp}$ is the exposure time.
Given that the total flux of a circular extended source is $F_{\rm obs}\sim A_{\rm o}\,\pi\sigma^2$ where $\sigma^2=(R_{1/2}^2+R_{\rm PSF}^2)/1.17^2$,  
   the SNR in the central pixel (i.e. the central SB$_c$, or $A_{\rm o}$) will scale as
\begin{eqnarray}
{\rm SNR}({A_{\rm o}})&\propto&\frac{A_{\rm o}\,t_{\rm exp}\,r_{\rm pix}^2}{\sqrt{{\rm SB}_{\rm sky}\,t_{\rm exp}\,r_{\rm pix}^2}}\nn\\
 &\propto&\frac{ \sqrt{t_{\rm exp}} }{ 
\left[R^2_{\rm PSF} + R_{\rm 1/2}^2\right]}{r_{\rm pix}}  \label{eq:SNR}
\end{eqnarray}
 where $R_{\rm PSF}$ the PSF radius, and $R_{1/2}$ the object half-light and $r_{\rm pix}$ the pixel size in arcseconds,
such that a change of 0\farcs2 in the PSF FWHM (from 0\farcs8 to 1\farcs0) corresponds to a fraction change of 15\%\ in S/N
for a galaxy of size $R_{1/2}=0\farcs6$, and accordingly
30\%\ more exposure time would be required to reach the same S/N.

\section{Conclusions}
\label{section:conclusions}

In this paper we presented an algorithm to constrain kinematic parameters of high-redshift disks directly  from 3-dimensional data cubes.
The algorithm uses a parametric model and  the knowledge of the 3-dimensional  kernel   to return a  3D modeled galaxy and a data cube
convolved with the 3D kernel. The parameters are estimated using an MCMC approach  with  nontraditional sampling distributions in order to efficiently probe the parameter space.

In summary,
\begin{enumerate}
\item the 2D version of the algorithm is used on an SDSS $r$-band image of a $z\sim0.2$ galaxy (Figure~\ref{fig:J1659})  taken at 1\farcs1 resolution. We find that the morphology is well recovered compared to a higher-resolution (0\farcs7)  CFHT image;
\item using a set of \Nideal\ mock data cubes,  Figure~\ref{fig:SBplot} shows that the accuracy on the recovered parameters depends on the product of the central SB,
SB$_{1/2, \rm obs}$ times the size-to-seeing ratio $(R_{1/2}/R_{\rm PSF})^{\sim1}$, following approximately the analytical expectation of \citet{RefregierA_12a};
\item from this set of mock data cubes, the morphological parameters do not depend on seeing, redshift, or the size-to-seeing ratio (Figure~\ref{fig:matrix_size});
\item from this set of mock data cubes, the robustness of the algorithm in recovering the kinematics parameters is also independent of seeing and redshift,
provided that the ratio between the galaxy half-light radius and the PSF radius $(R_{1/2}/R_{\rm PSF})$ is larger than 1.5 (Figure~\ref{fig:matrix_Vmax});
\item we also find that the accuracy in the recovered parameters does not depend on the FWHM accuracy, but depends more critically on the shape of the PSF,
except for the disk dispersion $\sigma_o$, which depends critically on the instrument LSF;
\item using a simulated disk galaxy from the hydro-simulation of Michel-Dansec et al., which contains asymmetric deviations, we found that
the kinematic parameters can be well estimated irrespectively of seeing, provided that the SNR is above a critical value
 (3 in this case; Figure~\ref{fig:global}).
Consequently, when the PSF FWHM is slightly above the original scientific goal (1\farcs0 instead of 0\farcs8) the optimal strategy is to integrate 30\%\ longer
(Equation~\ref{eq:SNR}) for a galaxy of size $R_{1/2}=0\farcs6$. 
\end{enumerate}
 
In conclusion, the \galpak\ algorithm can provide reliable constraints on  galaxy size, inclination, and kinematics over a wide range of seeing and of S/N. However, the algorithm should not be used blindly, and we stress that users of \galpak\ are strongly advised   (1) to look at the convergence of the parameters
(as in Figure~\ref{fig:mcmc});   (2) to investigate possible covariance in the parameters (as in Figure~\ref{fig:correl}), as these are rather data specific; and (3) to adjust the MCMC algorithm to ensure an acceptance rate between 30\%\ and 50\%, as discussed in the online documentation~\footnote{\url{http://galpak.irap.omp.eu/doc/overview.html}}.

Recent applications of the  \galpak\ algorithm can be found in \citet{PerouxC_13a,BoucheN_13a,SchroetterI_15a}, and \citet{BolattoA_15a}, which  illustrate the potential in using a global 3D fitting technique.

\acknowledgments
We are very grateful to the referee for his/her careful read of the manuscript and the detailed report that led to a significantly improved manuscript.
L.M.D. acknowledges support from the Lyon Institute of Origins under grant ANR-10-LABX-66.
Numerical simulations used in this work were performed using HPC resources from GENCI-CINES
(Grant 2013-x2013046642).  N.B. acknowledges support from a  Carreer Integration Grant (CIG) 
 (PCIG11-GA-2012-321702) within the 7th European Community Framework Program.
We warmly thank Antoine Goutenoir for his assistance in the current implementation and web services.



\begin{appendix}

\section*{Surface Brightnesses}

For extended sources with total flux $F_{\rm tot}$ and exponential profiles, i.e. SB$(r) \equiv I(r)$, one can define several measures of SBs. We have the following:
\begin{enumerate}
\item  The central SB $I_{\rm o}$ which is
	\begin{equation}
	I_{\rm o} = \frac{F_{\rm tot}}{2\pi R_{\rm d}^2} \label{eq:Io:exp}
	\end{equation}
in the case of an exponential flux profile $I(r)=I_{\rm o} \exp(-r/R_{\rm d})$   	{since $F_{\rm tot} = I_{\rm o} 2 \pi R_{\rm d}^2$},
where the half-light radius $R_{1/2}=1.68\;R_{\rm d}$.
In the case of a Gaussian flux profile  $I(r)=I_{\rm o} \exp(-r^2/2\sigma^2)$, it is
	\begin{equation}
	I_{\rm o} = \frac{F_{\rm tot}}{2\pi \sigma^2}  \label{eq:Io:gau}
	\end{equation}
  where the half-light radius $R_{1/2}=1.17\sigma$. 
\item  The average SB within the half-light SB$_{1/2}$:
	\begin{equation}
	{\rm SB}_{1/2} =  \frac{0.5\; F_{\rm tot}}{\pi R_{1/2}^2}\propto I_{\rm o}, \label{eq:rhalfSB}
	\end{equation}
	 where $R_{1/2}$ is the true or intrinsic  half-light radius ($R_{1/2}=1.68\;R_{\rm d}$).
\item The central pixel SB, ${\rm SB}_c$:
	\begin{eqnarray}
	{\rm SB}_c = A_{\rm o} \label{eq:SBc}
	\end{eqnarray}
	where the observed SB profile $F_{\rm obs}(r)$ is the convolution of $I(r)$ with the PSF $G(r)$, i.e.
	$F_{\rm obs}(r)\simeq A_{\rm o}\;\exp(-r^2/2\sigma^2)$ where now $\sigma$ contains the contributions from the intrinsic profile and
from the PSF via $(1.17\sigma)^2=R_{1/2}^2+R^2_{\rm PSF}$ ($=R^2_{1/2,\rm conv}$).
   $ R_{\rm PSF}$ is  the radius of the PSF ($R_{\rm PSF}\equiv$ FWHM/2).
\item The apparent central SB within the  half-light radius $R_{1/2,\rm conv}$, ${\rm SB}_{1/2, \rm conv}$:
	\begin{eqnarray}	
	{\rm SB}_{1/2, \rm conv} &=& \frac{0.5\; F_{\rm tot}}{\pi R_{1/2,\rm conv}^2} \simeq \frac{0.5\; F_{\rm tot}}{\pi (R_{1/2}^2+R_{\rm PSF}^2)}\label{eq:rhalfSBconv}
	\end{eqnarray}
	where  $R_{1/2,\rm conv}$ is the convolved half-light radius.
\item the observed central surface brightness  within the observed galaxy surface area $S_{1/2,\rm obs}$, ${\rm SB}_{1/2, \rm obs}$:
	\begin{eqnarray}
	{\rm SB}_{1/2, \rm obs} &=&  \frac{0.5\; F_{\rm tot}}{S_{1/2,\rm obs}} = \frac{0.5\; F_{\rm tot}}{\pi a b}\label{eq:Sconv}
	\end{eqnarray}
where $a$ and $b$ are the observed major and minor axis, respectively.
\end{enumerate}
The first three (Equations~\ref{eq:Io:exp}--\ref{eq:rhalfSB}) are not observable but can be derived
from the total flux $F_{\rm tot}$ and from the galaxy' s intrinsic size $R_{\rm d}$ or $R_{1/2}$.  On the other hand, 
the other two  (Equations~\ref{eq:SBc}--\ref{eq:rhalfSBconv}) are directly observable.

Naturally, the galaxy apparent area  $S_{1/2,\rm obs}$ is $\pi a\,b$ or $\pi a^2 \;(b/a)$; thus,
the face-on SB$_{1/2,\rm conv}$ (Equation~\ref{eq:rhalfSBconv}) and observed  SB$_{1/2,\rm obs}$ (Equation~\ref{eq:Sconv}) are related to one another 
 via the axis ratio $b/a$.\footnote{Generally speaking, 
$a\equiv R_{1/2,\rm conv} \simeq (R_{1/2}^2+R_{\rm PSF}^2)^{0.5}$ and 
$b \simeq (R_{1/2}^2 \cos^2(i)+R_{\rm PSF}^2)^{0.5}$.}

From these definitions, we now derive relationships between these variants of SB and begin by noting
that, typically for intermediate galaxies, the seeing radius $R_{\rm PSF}$ and the galaxy half-light radius $R_{1/2}$ are of the same order, i.e. $R_{\rm PSF}/R_{1/2}\sim1$.  Hence,
one can write $R_{\rm PSF}/R_{1/2} \sim (1-x)$ with $x\equiv (R_{1/2}-R_{\rm PSF})/R_{1/2}$ and $|x|<<1$. 

  Since the total flux $F_{\rm tot}=2\pi\;\sigma^2\;A_{\rm o}$ is also $2\pi\;R_{\rm d}^2\;I_{\rm o}$, we have 
\begin{eqnarray}
2\pi\;R_{\rm d}^2\;I_{\rm o} &=&A_{\rm o}\;2\pi\;\frac{R_{1/2}^2+R_{\rm PSF}^2}{2\ln(2)}\nn\\ 	
I_{\rm o} &\simeq&A_{\rm o} \;\frac{1.68^2}{\ln(2)}\;\frac{R_{\rm PSF}}{R_{1/2}},
\end{eqnarray}
which relates the observed S/N in the central pixel $A_{\rm o}$ to the intrinsic central SB $I_{\rm o}$.

  The average central surface brightness ${\rm SB}_{1/2, \rm conv}$ within $R_{1/2,\rm conv}$ is
\begin{eqnarray}
{\rm SB}_{1/2, \rm conv} &=&  \frac{0.5\; F_{\rm tot}}{\pi R_{1/2}^2}\frac{1}{ 1+(R_{\rm PSF}/R_{1/2})^2 }\nn\\
&\simeq& \frac{0.5\; F_{\rm tot}}{\pi R_{1/2}^2} \frac{1}{2}(1+x)\nn\\
&\simeq& 0.25\; {\rm SB}_{1/2} \times \frac{R_{1/2}}{R_{\rm PSF}}\propto  I_{\rm o}\times \frac{R_{1/2}}{R_{\rm PSF}}\label{eq:appendix}\\
&\simeq& 0.5 \frac{A_{\rm o}}{\ln(2)}\nn
\end{eqnarray}
which shows that the observed central SB, ${\rm SB}_{1/2, \rm obs}$,  directly maps onto the S/N in the central pixel.

\end{appendix}

\end{document}